\documentclass{article}

\usepackage{PRIMEarxiv}

\usepackage[utf8]{inputenc} % allow utf-8 input
\usepackage[T1]{fontenc}    % use 8-bit T1 fonts
\usepackage{hyperref}       % hyperlinks
\usepackage{url}            % simple URL typesetting
\usepackage{booktabs}       % professional-quality tables
\usepackage{amsfonts}       % blackboard math symbols
\usepackage{nicefrac}       % compact symbols for 1/2, etc.
\usepackage{microtype}      % microtypography
\usepackage{lipsum}
\usepackage{fancyhdr}       % header
\usepackage{graphicx} 
\usepackage{amsmath}% graphics
\graphicspath{{media/}}     % organize your images and other figures under media/ folder

\title{Network Dynamics in Mixed Martial Arts: A Complex Systems Approach to Ultimate Fighting Championship (UFC) Competition Insights
\thanks{\textit{msc796@alumnos.ucn.cl}} 
}

\author{
  Maximiliano S. Castillo \\
  Departamento de F\'{\i}sica, Universidad Cat\'{o}lica del Norte \\
  Antofagasta, Chile\\
  \texttt{msc796@alumnos.ucn.cl} \\
   \And
  Gabriel Fraczinet G.\\
  Facultad de F\'{\i}sica, Pontificia Universidad Cat\'{o}lica de Chile\\
  Santiago, Chile\\
  \texttt{gabriel.fraczinet@ug.uchile.cl} \\
   \And
  Víctor Muñoz\\
  Departamento de Física, Facultad de Ciencias, Universidad de Chile\\
  Santiago, Chile\\
  \texttt{vmunoz@macul.ciencias.uchile.cl} \\
}

\begin{document}
\maketitle

\begin{abstract}
The Ultimate Fighting Championship (UFC) has grown from a niche combat sport promotion into a globally recognized competitive enterprise. This study applies complex network analysis to explore the structural evolution of UFC matchmaking and its impact on competitive dynamics, fighter prominence, and audience engagement. By constructing directed and undirected networks where fighters represent nodes and bouts define edges, we examine key metrics such as degree distribution, clustering, betweenness centrality, and eigenvector centrality. Our findings reveal how the UFC’s matchmaking strategies transitioned from tightly clustered, repetitive matchups in its early years to a more decentralized and strategically curated fight network. We identify distinct structural properties between winners and losers, showing that successful fighters maintain centrality while frequently losing fighters exhibit surprising degrees of sustained connectivity. Correlations with Pay-Per-View sales and Google search trends suggest that network dispersion and novelty in matchups drive greater audience interest, while excessive clustering and density reduce engagement. Furthermore, comparisons with official rankings (Pound-for-Pound, champions, and top-15 fighters) demonstrate that traditional success metrics only partially align with network-based prominence, highlighting the complex interplay between structural connectivity, commercial appeal, and competitive success. This research contributes to the understanding of sports as complex adaptive systems and provides insights into how strategic matchmaking shapes both competitive integrity and economic viability in professional mixed martial arts.
\end{abstract}

% keywords can be removed

\section{Introduction}\label{sec:introduction}

\begin{figure}[!h]
    \centering
    \includegraphics[width = 0.75\textwidth]{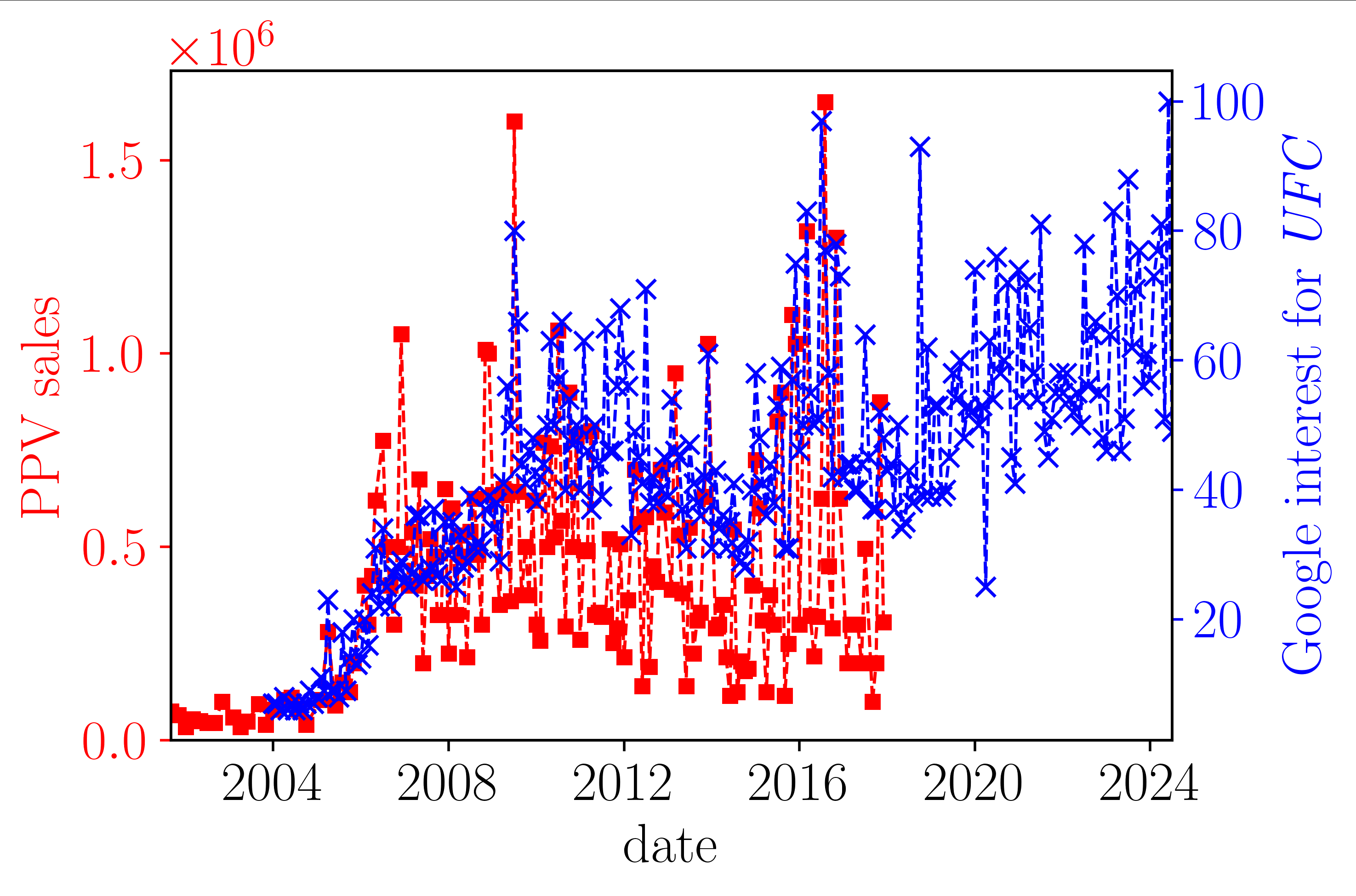}
    \caption{(red squares) UFC PPV sales. (blue crosses) Google interest for \textit{UFC} search.}
    \label{fig:interest}
\end{figure}

Mixed martial arts (MMA) has grown from a niche endeavor to a globally recognized sport, with the Ultimate Fighting Championship (UFC) at its forefront as both a cultural phenomenon and a billion-dollar enterprise~\cite{Robbins2017}. Economically, the UFC’s success is evident in its high Pay-Per-View (PPV) revenues (see Fig.~\ref{fig:interest}), lucrative broadcasting contracts, and widespread sponsorship deals. Sociologically, its influence extends well beyond athletic competition; it shapes fan communities, media discourse, and even national conversations on health, identity, and sporting ethics~\cite{Jennings2021MixedUFC, Snowden2010}. Viewed through this lens, the UFC’s rise is not merely about two fighters in a cage but also about how modern sports can transform into globally consumed spectacles that merge martial traditions, corporate strategies, and cultural narratives.

\begin{figure}[!h]
    \centering
    \includegraphics[width=0.75\linewidth]{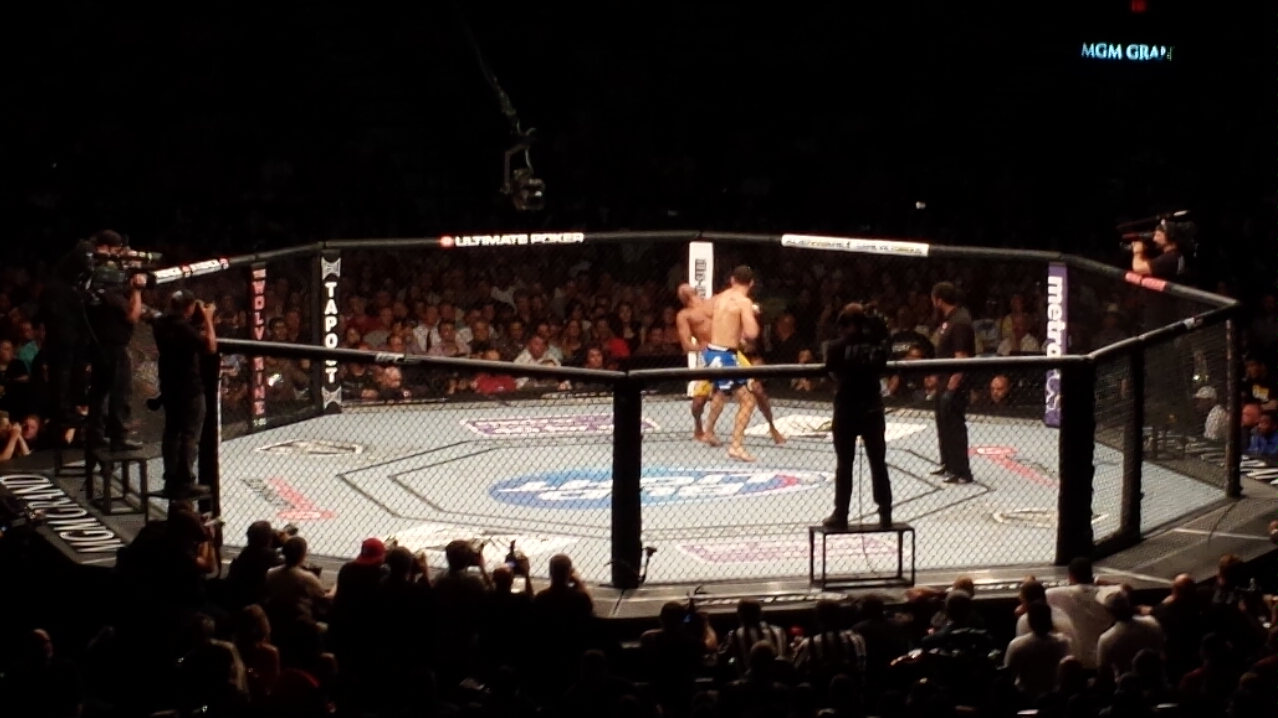}
    \caption{Chris Weidman knock out Anderson Silva at UFC 162. Extracted from Ref.~\cite{Kevlar2013}. Public domain license.}
    \label{fig:kevlar}
\end{figure}

Historically, the UFC traces its origins to November 12, 1993, when it held its inaugural event—UFC 1—at McNichols Sports Arena in Denver, Colorado. Created largely through the collaborative efforts of entrepreneur Art Davie, Brazilian jiu-jitsu co-founder Rorion Gracie, and television executive Bob Meyrowitz, the initial concept stemmed from the desire to pit diverse combat styles against each other in an unscripted format \cite{Davie2014, Gentry2011}. This first event was promoted as the answer to a seemingly age-old question: “Which fighting style is truly superior?”

In these formative years, the UFC operated under minimal rules, reflecting its raw, experimental character. Fighters wore no mandatory gloves, weight classes were non-existent, and events often featured single-night tournaments in which athletes competed multiple times until a sole victor emerged. The open-ended rule set—often referred to as “no holds barred” (NHB)—made for an unregulated, and at times controversial, competition environment. Early champions like Royce Gracie showcased Brazilian jiu-jitsu’s dominance over bigger, stronger adversaries, while other notable figures such as Ken Shamrock and Kimo Leopoldo became key attractions by blending wrestling, catch submission techniques, and traditional martial arts \cite{Gentry2011, Jennings2021MixedUFC}.

These early events, though influential in generating curiosity about mixed martial arts, also faced broad criticism and political pushback. Senator John McCain famously labeled the emerging sport as “human cockfighting,” prompting many cable systems to drop the pay-per-view broadcasts \cite{Schildhauer2013}. Despite such controversies, this initial “rule-defying” era laid the foundation for the modern UFC, igniting fan interest in cross-disciplinary combat while foreshadowing the regulatory changes and professionalization. 

Over the ensuing decades, the Ultimate Fighting Championship (UFC) underwent a series of major transformations that significantly altered both its structure and public profile. The most pivotal change came in January 2001, when the Fertitta brothers (Lorenzo and Frank Fertitta III) and Dana White purchased the UFC’s parent company—then known as SEG (Semaphore Entertainment Group)—for a reported 2 million, subsequently forming Zuffa, LLC \cite{Snowden2010}. This transition, often referred to as the Zuffa era, ushered in sweeping reforms aimed at legitimizing mixed martial arts (MMA) and bringing the UFC into mainstream acceptance. Structured Weight Classes and Unified Rules
Shortly after Zuffa’s acquisition, the UFC embraced the Unified Rules of MMA, which had been codified in part by the New Jersey State Athletic Control Board. These rules introduced formalized weight classes, including lightweight (155 lbs), welterweight (170 lbs), middleweight (185 lbs), and light heavyweight (205 lbs), among others. They also mandated the use of gloves, restricted illegal moves (e.g., head-butting, groin strikes), and clearly defined round lengths and cage regulations \cite{Schildhauer2013}. This framework was crucial in combating prior criticism regarding the sport’s perceived lack of safety and paved the way for broader athletic commission oversight across the United States.

Innovations in Fighter Safety and Matchmaking
Under Zuffa, the UFC ramped up fighter medical screenings, introduced pre- and post-fight drug testing, and established policies around weight-cutting protocols to enhance athlete health \cite{Jennings2021MixedUFC}. In parallel, newly appointed matchmakers helped develop a rigorous matchmaking system that balanced competitive parity with fan-friendly bouts, thereby improving both the fairness and entertainment value of events. These reforms not only reduced injuries but also fostered a sense of professional credibility that encouraged more high-level athletes from disciplines such as wrestling, Muay Thai, and jiu-jitsu to enter the UFC ranks.

In its current form, the UFC has secured its place as the world’s premier mixed martial arts organization, boasting a global roster that exceeds 600 active fighters across multiple men’s and women’s weight divisions. Post-2016 ownership changes (when WME-IMG—now Endeavor—purchased the UFC for 4.025 billion dollars) have not slowed its expansion; rather, the promotion continues to host more than 40 events per year on nearly every continent (see Fig.~\ref{fig:interest}). Economically, the UFC now stands as one of the most valuable properties in combat sports, frequently surpassing one million PPV buys for marquee matchups and leveraging sponsorships, merchandising, and digital streaming platforms to reach a growing global audience \cite{Merced2016}. These milestones reflect a broader transition from niche “no holds barred” spectacles toward a structured and economically robust global sport.

Such ongoing growth and complexity make the UFC an illustrative subject for complex systems analysis. In complex systems research, interactions among individual components (e.g., fighters, matchmakers, athletic commissions) can produce emergent phenomena—collective behaviors and patterns not fully predictable by examining only the components themselves \cite{Newman2003, Barabasi2002b}. Within the UFC, fighter matchups form a dynamic, evolving network in which each node (fighter) is linked to others through competition, and outcomes feed back into fan interest, promotional decisions, and fighter trajectories. Network science methods can unravel how hierarchies form, how “hub” fighters command disproportionate attention, and how periodic reorganizations (such as the introduction of new weight classes or regulations) reshape the competitive landscape \cite{Radicci2011}. Moreover, analyzing the UFC through a complex systems lens facilitates the identification of tipping points—periods when structural changes in matchmaking or global branding lead to surges in PPV sales or media visibility. As such, the UFC’s transformation over the past two decades provides both a case study of global sports commercialization and a fertile domain for applying models and metrics from the interdisciplinary field of complex systems research.

Complex network theory offers a powerful framework for analyzing systems of both technological and scholarly relevance \cite{Barabasi2002}. Illustrative applications range from modeling the chemical reaction networks that underpin cellular function \cite{Stein2009, Cherubini2015, Albert2005b, Costa2008} and investigating the intricacies of protein folding \cite{Braun2012, Spirin2003, Tucker2001, Araujo2007}, to examining tectonic plate interactions and seismic processes \cite{Pasten2018, Pasten2021, Pasten2018b, Pavez2023, Baiesi2005, Abe2006, Jimenez2013}, exploring complex computational problems \cite{Nitta2003, Pasqualetti2014, Hanrahan2010}, and studying social structures \cite{Braha2009, Jalili2013, Klemm2003}. At its core, this approach reduces multifaceted phenomena to fundamental constituents (nodes) and their interconnections (edges), thereby elucidating the underlying principles and emergent properties that govern a wide variety of complex systems.

In the present analysis, we will regard fighters “nodes,” while their matches form “edges”—a representation that allows for rigorous, mathematically grounded analyses of how fighters interconnect over time. This study draws upon UFC match data spanning from 1994 to 2021, capturing a transformative period characterized by transitions in ownership, rules, and roster size. We construct both directed and undirected networks to capture different dimensions of competition: a fighter beating another introduces direction (winner to loser), while an undirected tie simply indicates that two athletes have fought. Additionally, we differentiate between networks built solely on winners, solely on losers, and on traditional matchmaking to glean multiple perspectives on how success and defeat shape the underlying topology.

We then apply a suite of statistical and structural metrics that elucidate the network’s evolving properties. Measures such as degree, betweenness centrality, and eigenvector centrality highlight how fighters with multiple connections or strategic “bridge” positions can stand out within the competitive fabric. We also assess path lengths, clustering coefficients, and overall network density to capture changes in how “clustered” or “diffuse” the sport’s matchmaking became as it expanded across weight classes and continents. Crucially, we correlate these network measures with PPV sales and Google search data—external metrics that gauge fan interest and commercial success. This dual approach offers insight into whether, for instance, having a more interconnected roster or a handful of central star fighters is linked to higher audience engagement.

Our findings trace how, in the early UFC, high connectivity and repeated matchups among a few combatants led to a tightly knit network structure, reflecting single-night tournaments and small fighter pools. As the organization introduced new rules, expanded weight divisions, and curated diverse rosters, the network became less dense and more distributed. Fighters who consistently win maintain a greater number of influential ties—reinforcing their visibility—while those with frequent losses occupy a more peripheral or transient position. Yet intriguingly, being a “high-degree” loser can still correlate with fan interest if a fighter’s style or persona resonates with audiences (as seen in several high-profile “fan-favorite” fighters who have not always enjoyed winning records).

On a larger scale, these network metrics align with the UFC’s commercial trajectory. Periods where the network offered fresh matchups and larger rosters often coincided with spikes in PPV numbers and Google searches, suggesting that novelty and diversity in the sport’s matchmaking are compelling draws. Conversely, clustering and repetitive fights tended to diminish that interest. Furthermore, while some network measures correlated well with top-15 divisional rankings—indicating that active fighters who face many opponents remain competitively relevant—there was a weaker link to the more subjective pound-for-pound (P4P) lists. These observations underscore the multifaceted nature of fighter “success,” which fuses competitive performance, promotional narratives, and fans’ thirst for novelty.

In sum, by treating the UFC’s history as a dynamic network of interlinked fights and fighters, we capture both the sport’s structural evolution and its real-world economic and sociological footprints. In this first approximation, the analysis demonstrates that connectivity and dispersion in the roster significantly affect fan interest, highlighting how strategic matchmaking can sustain both competitive integrity and commercial appeal. Nevertheless, the story of the UFC is also one of individual charisma, organizational strategy, and shifting cultural norms—forces that pure network metrics cannot fully encapsulate. This study thus provides a comprehensive picture of how structural dynamics and consumer engagement interplay in one of the fastest-growing sports arenas, while also paving the way for future, more granular analyses that may include factors like social media presence, fighter personalities, and the role of cross-promotional endeavors.

\section{Methods}

\subsection{The UFC data}\label{sec:data}

The dataset under examination comprises historical UFC matches spanning from \textit{UFC 2}, which transpired on March 11, 1994, to \textit{UFC on ESPN: Brunson vs. Holland}, held on March 20, 2021, and can be downloaded from Ref.~\cite{Warrier2021}. The documents includes the date of the match, the winner, the method of victory, the venue, and supplementary details regarding attacks and grappling techniques.

A comparative analysis is undertaken between the network metrics, Pay-Per-View (PPV) sales and Google search interest in the term \textit{UFC} and its roster, serving as a metric for event popularity. The datasets, available from Refs.~\cite{Tapology2024} and \cite{Google2024} respectively, provide a quantitative assessment of PPV sales per paid event and the overall popularity of the sport and its fighters.

\subsection{Studied metrics}\label{sec:methodMetrics}

In this study, we used a variety of metrics to analyze and describe the structural properties of the networks under consideration. One of the fundamental metric is the \textit{degree} \(k\), which measures the number of connections a node has. This provides an essential measure of connectivity in the network. We also examined the \textit{clustering coefficient} $C$, which quantifies the tendency of a node’s neighbors to form connections with each other, revealing local structural patterns. This metric is defined as

\begin{equation}
C(\nu) = \frac{2T_{\nu}}{k_{\nu}(k_{\nu}-1)},
\end{equation}

where \(T_{\nu}\) is the number of triangles centered at node \(\nu\), and \(k_{\nu}\) is its degree \cite{Sizemore2018}. At a global level, we evaluated the \textit{density} $D$ of the network, a measure of how many connections exist compared to the total possible connections. It is given by

\begin{equation}
D = \frac{2E}{N(N-1)},
\end{equation}

where \(E\) is the number of edges, and \(N(N-1)\) the maximum number of edges in a fully connected network \cite{Rezwan2012}. The \textit{average path length} $l$ was also calculated. It measures the average shortest distance between pairs of nodes as

\begin{equation}
l = \sum_{i, j \in V}\frac{d(i, j)}{N(N-1)},
\end{equation}

where \(d(i, j)\) is the shortest path distance between nodes \(i\) and \(j\), and \(V\) represents all nodes~\cite{Zurita2023}. To assess the role of specific nodes in facilitating network connectivity, we used \textit{betweenness centrality} $b$. It measures how often a node lies on the shortest path between other nodes:

\begin{equation}
b(\nu) = \frac{1}{N} \sum_{i, j \in V} \frac{\sigma_{ij}(\nu)}{\sigma_{ij}},
\end{equation}

where \(\sigma_{ij}(\nu)\) is the count of shortest paths passing through node \(\nu\), and \(\sigma_{ij}\) is the total number of shortest paths between nodes \(i\) and \(j\)~\cite{Sizemore2018}. Finally, we employed \textit{eigenvector centrality} $\lambda$, a spectral metric that determines node importance based on its connections to other influential nodes. The principal eigenvector of the adjacency matrix \(A\) is calculated as:

\begin{equation}
Ax = \lambda x,
\end{equation}

where \(x\) is the eigenvector corresponding to the largest eigenvalue \(\lambda\). Metrics regarding nodes can be averaged for the whole network as follows
\begin{equation}
    \langle x\rangle = \frac{1}{N_\text{nodes}}\sum_i x_i,
\end{equation}
where $\langle x\rangle$ is the averaged value, $N_\text{nodes}$ is the number of nodes, and $x_i$ is the value of the metric for node $i$.

\subsection{Network construction}

In this work, we consider three different approaches for the construction of the networks, in an effort to understand the UFC system through diverse analytical lenses. Each approach considers a fighter as a node within the network, however, the definition of the edges will depend on the chosen methodology, with the anticipation that these edges will capture and signify distinct facets of the system. 

Our first approach is to construct the network based in the most simple competitive aspect. We established an edge on the sole condition that the fighters have had a match. In this regard, the information about the loser and winner can be included as the direction of the edge i.e. if fighter \textit{A} wins over fighter \textit{B}, a directed connection will be established from node \textit{A} to node \textit{B}. This approach is anticipated to provide insights into the inherent organizational dynamics of UFC fight arrangements.

Another methodology is based upon an extensive literature around non-linear dynamics and seismicity, and its formulation from a complex networks standpoint~\cite{Abe2004, Abe2006, Abe2011, Pasten2018}. In an analogous manner to these works, we construct two main networks, the \textit{winners network}, and the \textit{losers network}. The first one is constructed by connecting the node corresponding to a winner of a bout at $t_1$, to the winner of a bout at $t_2$. On the other hand, the losers network is constructed by connecting the node corresponding to the loser of a bout at $t_1$, to the lose of the bout at $t_2$.

\begin{figure}[!h]
    \centering
    \includegraphics[width=0.6\linewidth]{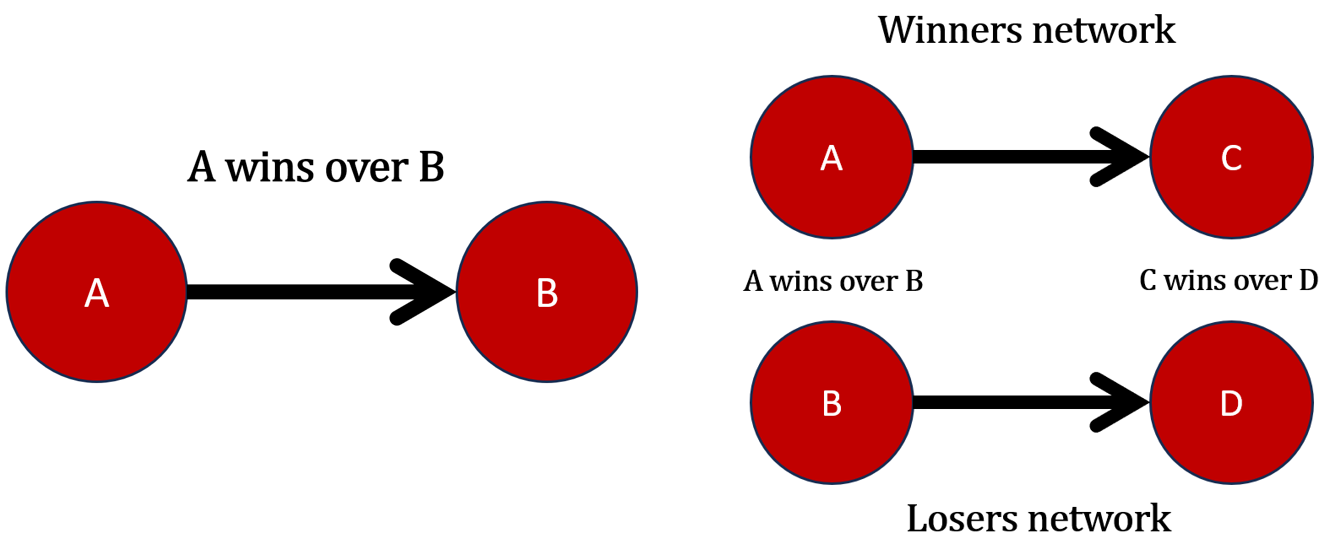}
    \caption{Methods for the networks construction. (left) matchmaking criteria: here the network is constructed based upon the winner and the loser of a bout. (right) winners/losers criteria: here two networks are constructed, the first one is constructed by connecting the winners of the bouts, while the later is constructed by connecting the losers, the edge direction stores the temporal information.}
    \label{fig:constructionMethod}
\end{figure}

To better understand the temporal evolution of the network topology, we will base our study in the fundamentals of dynamical networks~\cite{Zou2014}. For this extent, we will establish the connections in two-years temporal windows, with 1 month-delay. For each of these steps, we will calculate the metrics described in Sec.~\ref{sec:methodMetrics} for the resulting networks. 

\section{Results}

%---------------------------------------------
\subsection{Network metrics}\label{sec:degree}

\begin{figure}[h!] \centering \includegraphics[width=0.6\textwidth]{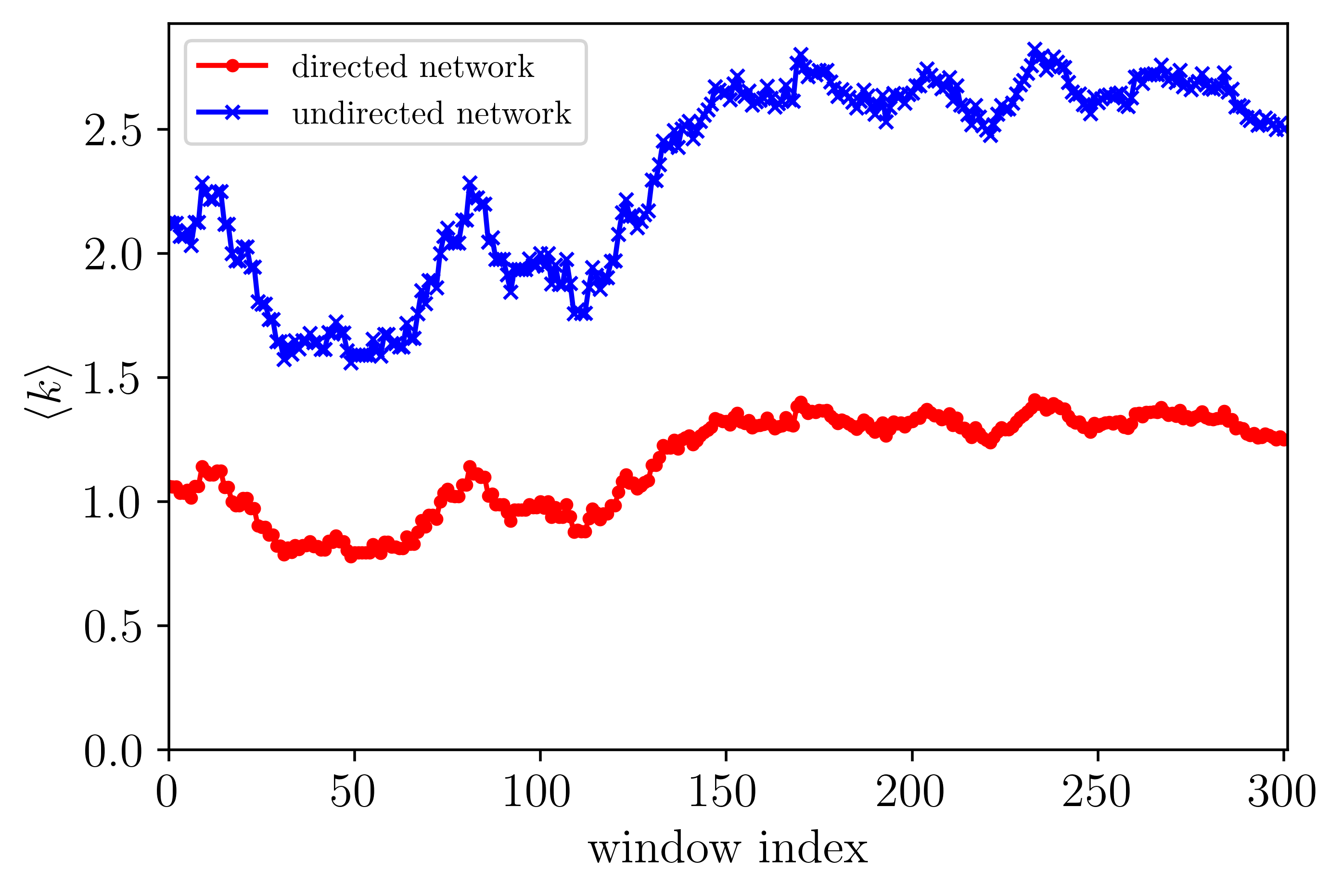} \caption{Average degree distribution $\langle k \rangle$ of the directed and undirected networks based on matchmaking criteria.} \label{fig:degreeMethod1} \end{figure}

\begin{figure}[h!]
    \centering
    \includegraphics[width = 0.7\textwidth]{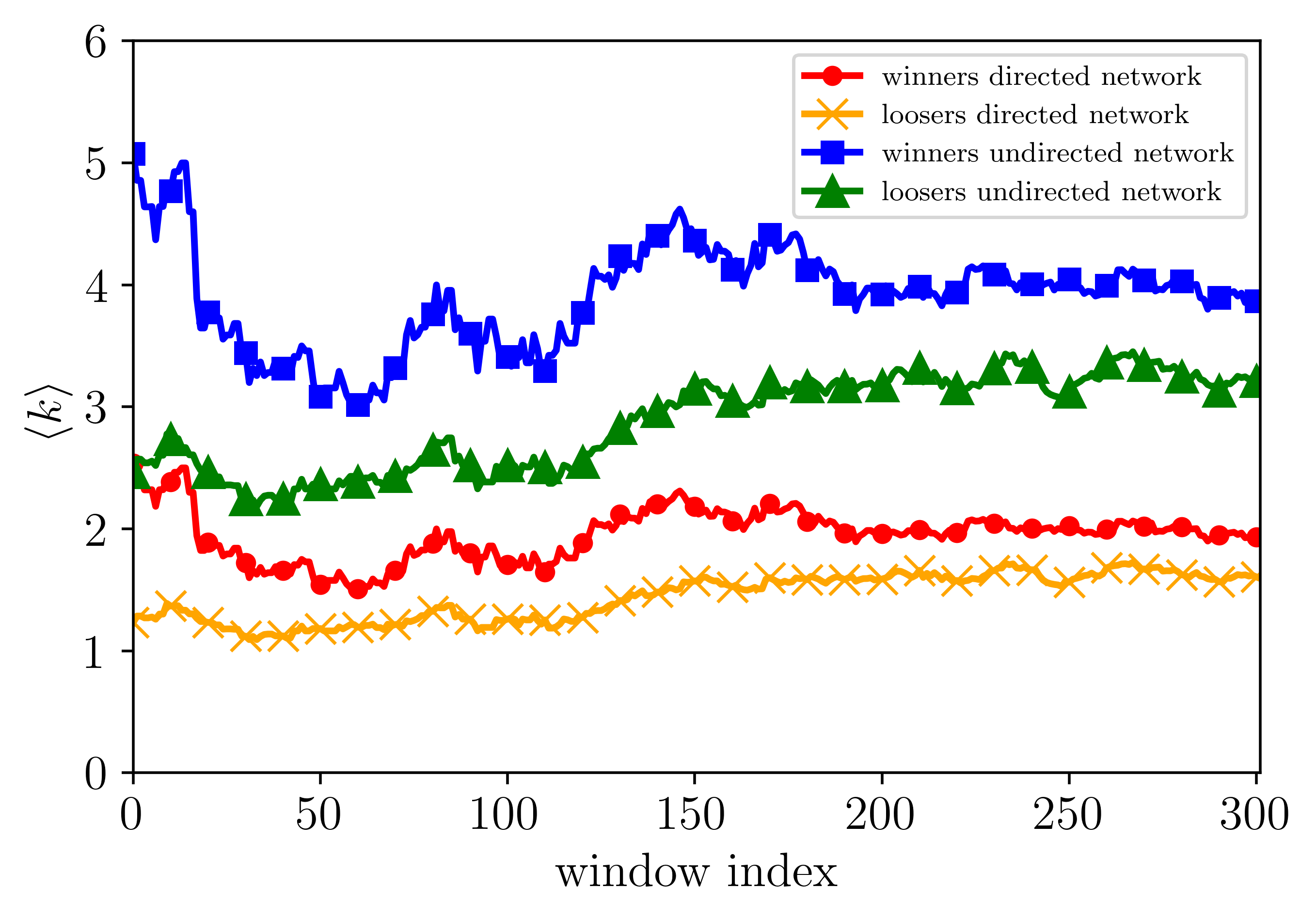}
    \caption{Average degree of the network constructed under the winners/losers criteria.}
    \label{fig:degreeMethod2}
\end{figure}

The evolution of the average degree $\langle k \rangle$ across the UFC fighter network provides a foundational understanding of its structural dynamics and competitive balance over time. Fig.~\ref{fig:degreeMethod1} and Fig.~\ref{fig:degreeMethod2} illustrate these trends for the directed and undirected networks based on matchmaking, and for the winners and losers networks respectively.

In the early years of the UFC, both directed and undirected matchmaking networks (Fig.\ref{fig:degreeMethod1}) exhibited relatively low average degrees, approximately $\sim 2$ and $\sim 1$, respectively. During this period, fighters frequently faced the same opponents multiple times, particularly within single-night tournaments, resulting in highly interconnected structures\cite{Jennings2021MixedUFC}. However, since these metrics are computed using two-year sliding windows, the expected concentration of connections is attenuated. Additionally, the brevity of fighters’ careers in the early UFC further contributed to this effect.

Directed networks, which emphasize the hierarchical nature of fight outcomes, consistently display lower average degrees. This pattern underscores the dominance of a select group of fighters, as victories tend to be concentrated within a small subset of competitors. Such structural characteristics align with observations in social and biological networks, where directed configurations often reveal concentrated power dynamics~\cite{Barabasi2002, Newman2003}.

As the window index progresses, a diminution of the average degree in both networks can be observed. This period correspond to the transition to the Zuffa era, when significant organizational changes, such as roster expansion and the introduction of new weight classes was implemented. By the window index of 100, $\langle k \rangle$ had dropped to approximately 1.75 in the undirected network and below 1.0 in the directed network. This decline underscores the organization’s efforts to diversify matchmaking and reduce the dominance of specific fighters. Then, a temporary peak is observed in both networks around this window index, reflecting a period of intensified competition marked by high-profile rivalries and an increase in events.

The winners/losers networks provide additional perspectives, as shown in Fig.~\ref{fig:degreeMethod2}. The undirected winners network exhibits the highest degrees, stabilizing at approximately 3.5 in later windows, reflecting the sustained prominence of successful fighters. By contrast, the undirected losers network stabilizes at lower values $\sim 2.5$, illustrating the more transient nature of losing fighters within the network. Directed winners and losers networks further emphasize this disparity, with winners maintaining higher degrees due to their frequent and strategic placement in central matchups. The structural advantage of successful fighters aligns with patterns observed in systems where dominant nodes maintain high connectivity, as seen in both ecological, geophysical and social networks~\cite{Rezwan2012, Abe2004, Pasten2017}.

When compared, the trends in the directed and undirected winners and losers networks highlight the contrasting dynamics of success and defeat. The higher average degree in the winners networks reflects the UFC’s emphasis on sustaining the prominence of victorious fighters, who often remain central to matchmaking strategies. In contrast, the lower connectivity of the losers networks captures the more transient and distributed nature of defeat, with losing fighters frequently cycling out of high-profile narratives. This division between winners and losers networks reveals the UFC’s dual strategy of promoting dominant fighters while maintaining a diverse and evolving roster.

In later temporal windows, beginning around window $\sim 80$, the stabilization of average degree values across all networks reflects the maturation of the UFC’s matchmaking strategies. The undirected networks maintain consistently higher average degrees, indicating a broad distribution of competitive interactions across the roster. In contrast, the directed networks emphasize the hierarchical structure of victories and losses, underscoring the consolidation of dominance among select fighters. Together, these network dynamics illustrate the UFC’s capacity to balance the promotion of dominant narratives with the necessity of maintaining competitive diversity. This equilibrium fosters both sustained audience engagement and a dynamic competitive environment, reinforcing the organization’s long-term strategic development.

%----------------------------------
%CLUSTERNG:
%-----------------------------------

\begin{figure}[h!] \centering \includegraphics[width=0.7\textwidth]{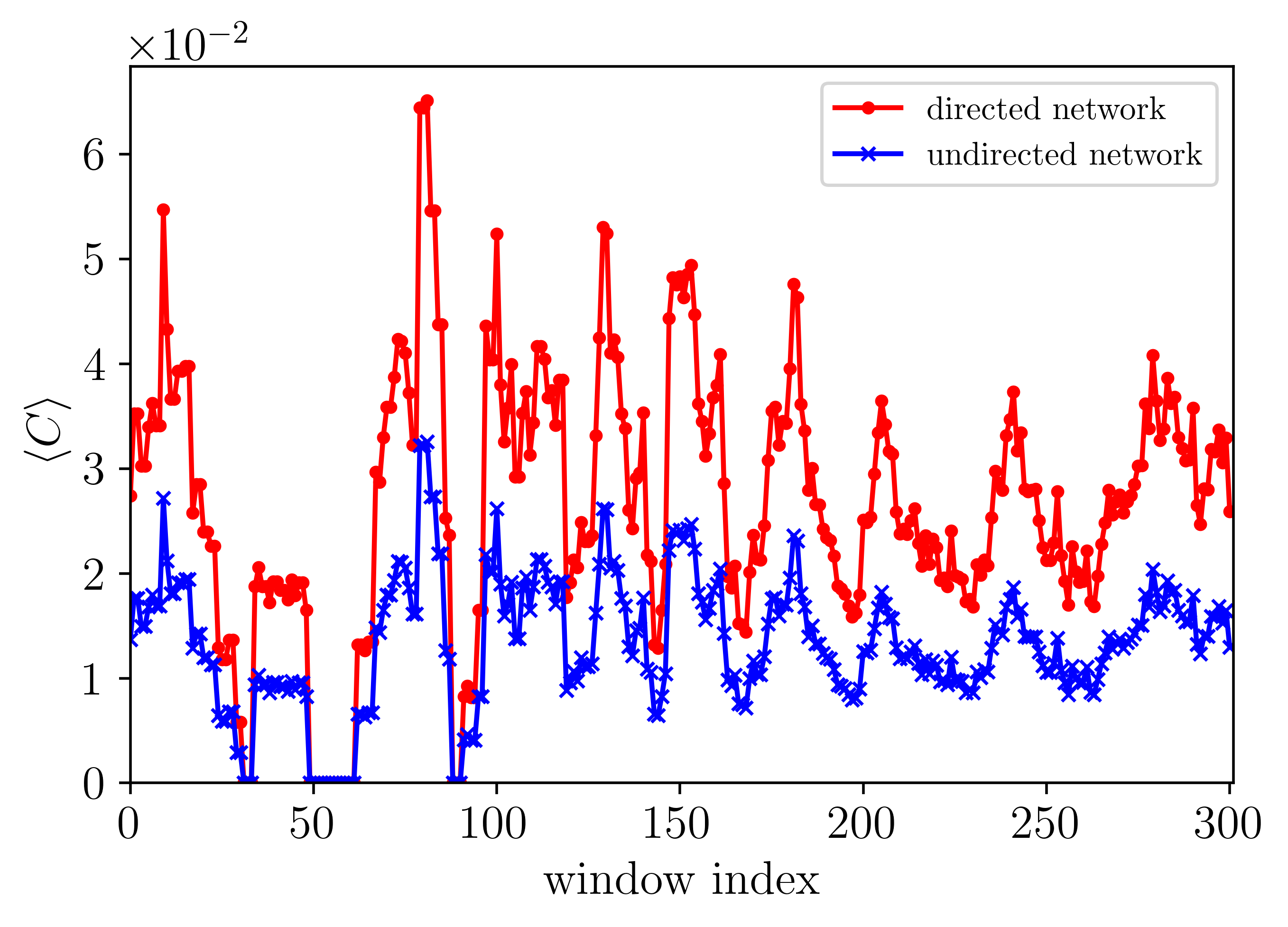} \caption{Clustering coefficient of the directed and undirected networks based on matchmaking.} \label{fig:clusteringMethod1} \end{figure}

\begin{figure}[h!]
    \centering
    \includegraphics[width = 0.7\textwidth]{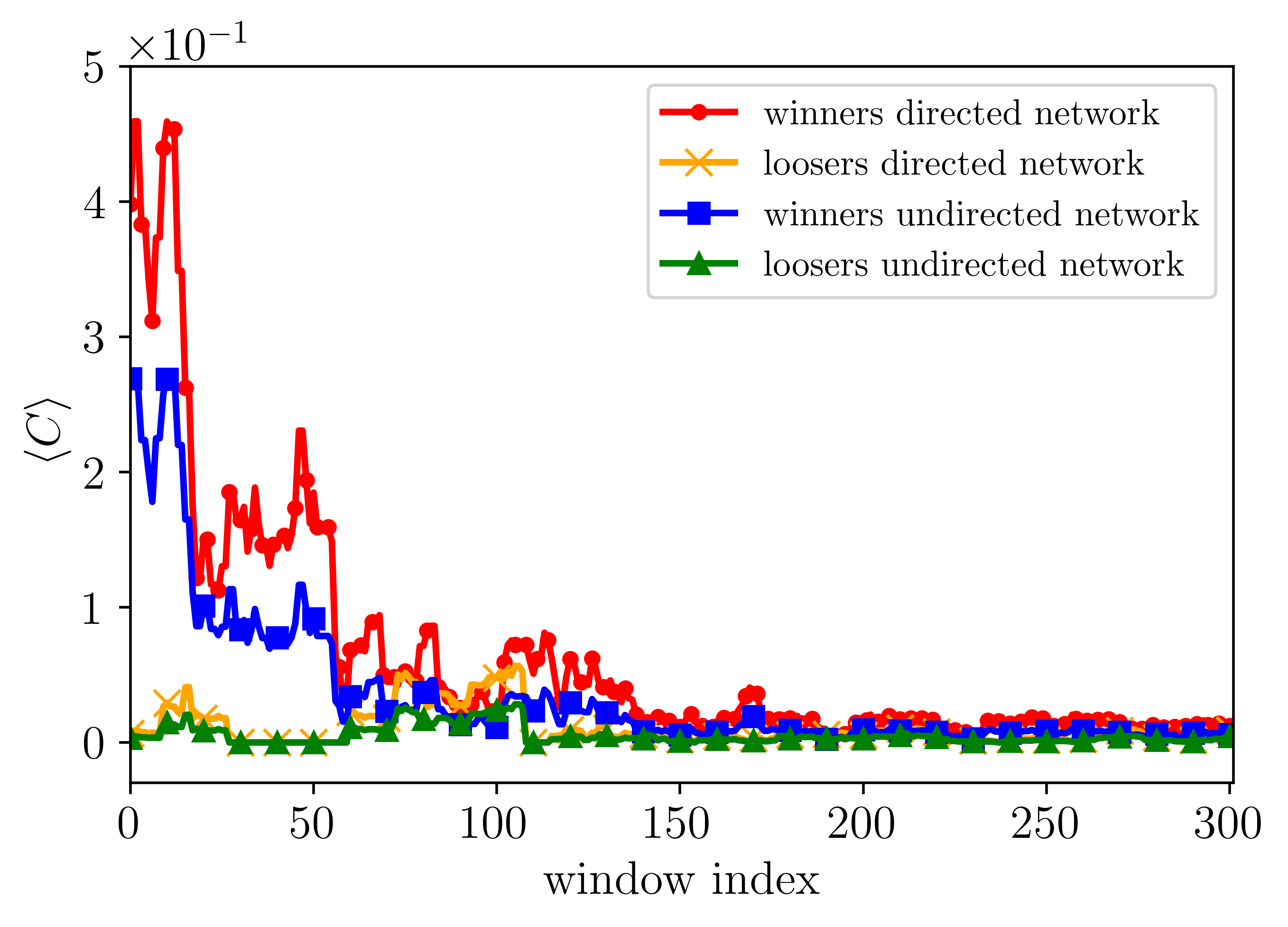}
    \caption{Clustering coefficient of the network constructed under method 2.}
    \label{fig:clusteringMethod2}
\end{figure}

In this context, the clustering coefficient $\langle C \rangle$ measures the degree to which fighters who have competed against each other also share common opponents, revealing the network's tendency to form tightly-knit groups. Fig.~\ref{fig:clusteringMethod1} and Fig.~\ref{fig:clusteringMethod2} illustrate these trends for both matchmaking and winners and losers networks.

The clustering coefficient is initially high for both directed and undirected matchmaking-based networks (Fig.~\ref{fig:clusteringMethod1}), at approximately $6\times10^{-2}$. This high value reflects the UFC’s early years, during which the small roster and repetitive matchmaking fostered dense connectivity and overlapping interactions. Fighters often faced the same opponents repeatedly, particularly within single-night tournaments, leading to tightly-knit clusters. Dominant figures like Royce Gracie and Don Frye are examples of these these trends, as their repeated success against multiple opponents reinforced hierarchical structures~\cite{Gentry2011, Jennings2021MixedUFC}. The directed network consistently exhibits higher clustering coefficients than the undirected network throughout the timeline. This difference underscores the directed network’s sensitivity to competitive hierarchies, as it emphasizes the outcomes of fights, creating stronger clusters around dominant fighters. In contrast, the undirected network captures broader connectivity without distinguishing between victories and losses, resulting in a more diluted clustering pattern.

Over time, the clustering coefficient in both matchmaking networks steadily declines, approaching near-zero values. Similarly to the results for the degree (see Sec.~\ref{sec:degree}), this decline aligns with the structural transformations initiated under the Zuffa administration, including roster expansion and diversified matchmaking. As the UFC introduced more fighters, the density of connections decreased, reducing the likelihood of shared opponents and dismantling the tightly-knit clusters characteristic of its early years. The directed network’s sharper decline highlights its responsiveness to these changes, capturing the redistribution of competitive outcomes across a broader pool of fighters. This redistribution reflects the UFC’s transition toward a more balanced structure, where dominance is less concentrated and victories are spread more evenly among competitors.

The winners and losers networks in Fig.~\ref{fig:clusteringMethod2} reveal additional nuances in clustering dynamics. The directed winners network exhibits the highest clustering coefficients across all configurations, reflecting the sustained connectivity and centrality of successful fighters. These fighters frequently compete in successive bouts, often sharing common opponents within tightly-knit groups. In contrast, the undirected and directed losers network shows consistently lower clustering coefficients, indicating a less interconnected structure among fighters who experience defeat. This disparity highlights the differing roles of winners and losers within the UFC’s competitive framework, with winners forming cohesive subgroups and losers participating in a more dispersed network.

%The directed winners and losers networks provide further insights into the hierarchical nature of these subgroups. The directed winners network initially exhibits a high clustering coefficient, reflecting the dominance of specific fighters who consistently win and share common opponents. Over time, this clustering declines but stabilizes at a level that emphasizes the persistence of localized hierarchies among successful fighters. Conversely, the directed losers network consistently exhibits the lowest clustering coefficients, reflecting the transient and decentralized nature of defeat in the UFC. Fighters who lose are less likely to participate in successive bouts against shared opponents, resulting in weaker clustering and more dispersed connectivity.

Interestingly, both matchmaking-based and winners/losers networks exhibit temporary peaks in clustering coefficients, particularly around the window index of 50 and again near index 100. These peaks likely correspond to periods of heightened activity within specific subgroups of fighters, driven by rivalries, trilogies, or narrative arcs that reinforce existing clusters. For example, rivalries such as Chuck Liddell versus Tito Ortiz created localized bursts of connectivity, as these fighters frequently faced each other and shared opponents during their prime years~\cite{Gentry2011, Jennings2021MixedUFC}. These peaks are more pronounced in the directed networks, which capture the hierarchical dynamics of these rivalries and their impact on clustering.

In the later temporal windows, the clustering coefficients stabilize at lower levels for all networks. The winners networks retain higher clustering coefficients compared to the losers networks, underscoring the sustained centrality of successful fighters within the UFC’s ecosystem. This stabilization reflects a mature and balanced structure, where tightly-knit clusters become rarer, and interactions are distributed across a larger and more diverse roster.

%The trends observed in these networks align closely with the UFC’s historical development. During its early years, high clustering coefficients reflect the organization’s small, centralized roster and repetitive matchmaking. As the UFC professionalized under Zuffa, the decline in clustering reflects the organization’s efforts to diversify competition, globalize its roster, and introduce new weight classes. The winners and losers networks further illustrate the contrasting dynamics of success and defeat, with winners forming more cohesive and enduring clusters, while losers contribute to a broader and more transient competitive structure.

%---------------------------------------
%DENSITY
%---------------------------------------
\subsubsection{Density}

\begin{figure}[h!] \centering \includegraphics[width=0.7\textwidth]{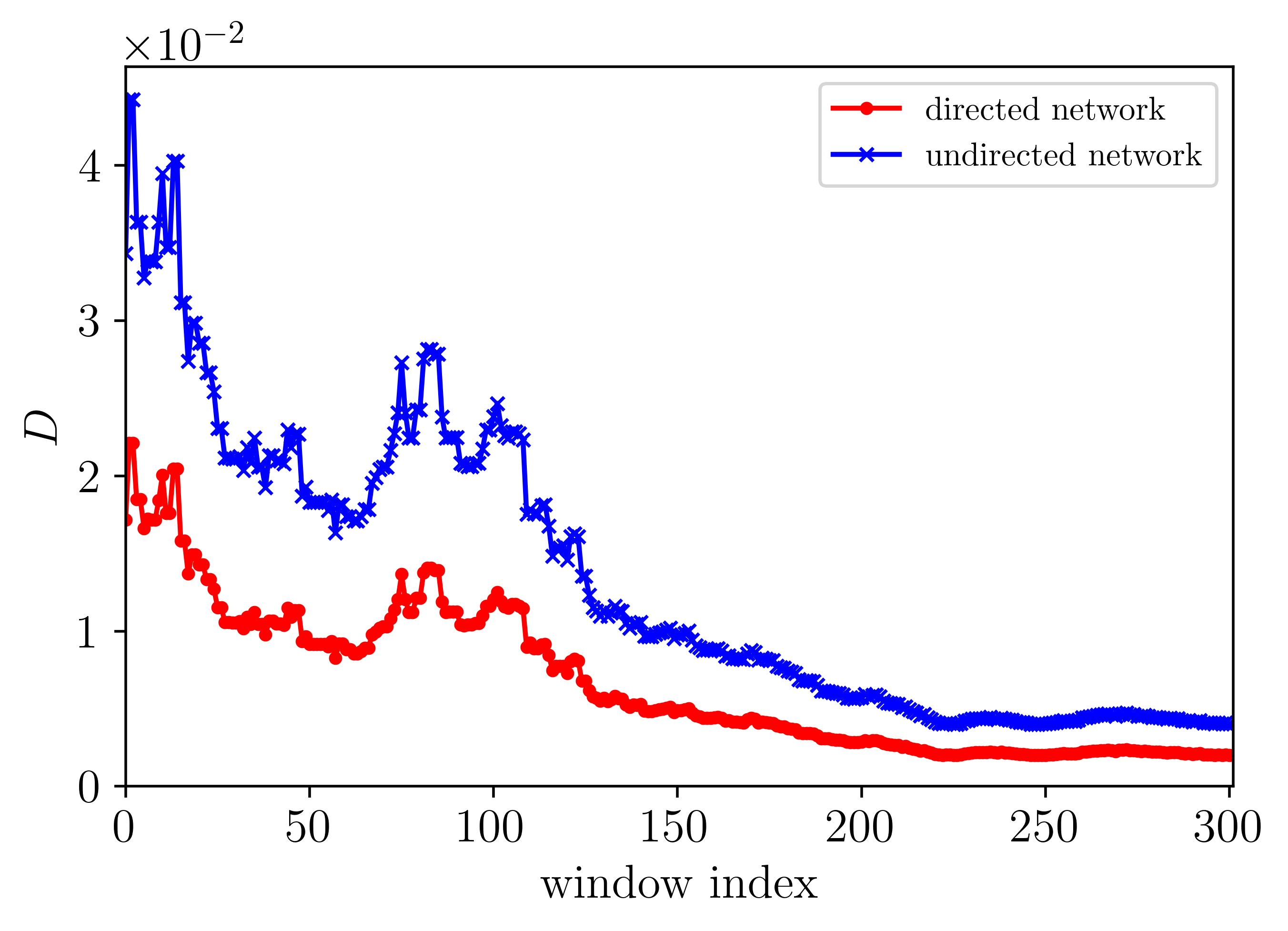} \caption{Density of the directed and undirected networks based on matchmaking.} \label{fig:normDegMethod1} \end{figure}

\begin{figure}[h!]
    \centering
    \includegraphics[width = 0.7\textwidth]{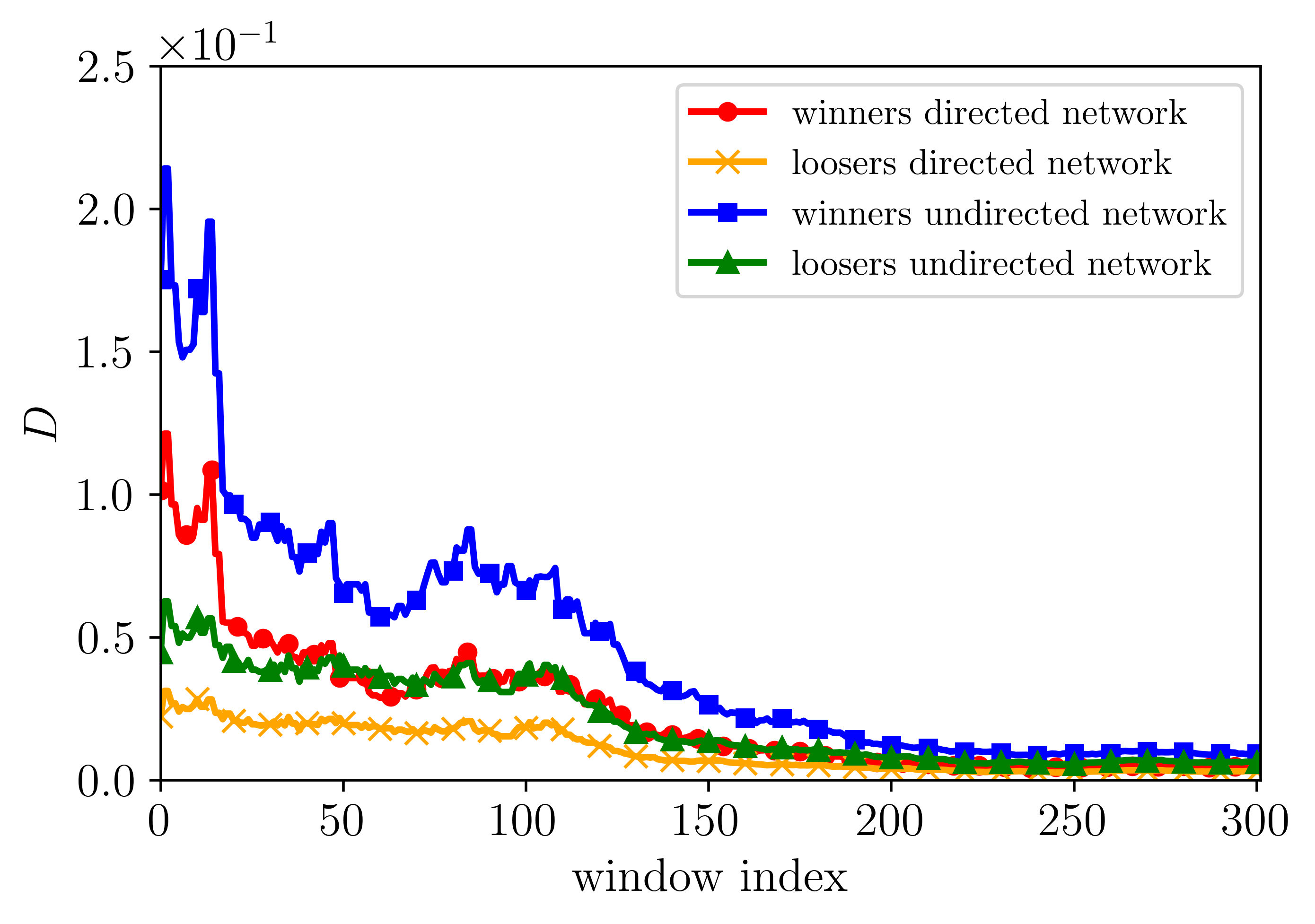}
    \caption{Density of the network constructed under method 2.}
    \label{fig:densityMethod2}
\end{figure}

The evolution of the network density $D$ of the UFC fighter network across temporal windows provides a detailed understanding of its structural development. Fig.~\ref{fig:normDegMethod1} and Fig.~\ref{fig:densityMethod2} illustrate the trends for the directed and undirected matchmaking-based networks and the winners and losers networks, respectively.

Initially, the network density is relatively high for both directed and undirected matchmaking-based networks (Fig.~\ref{fig:normDegMethod1}), starting at approximately $4.5\times10^{-2}$ for the undirected network and approximately half of this value for the directed network. This reflects a period when a substantial proportion of potential connections were realized, resulting in a highly interconnected network. As mentioned before, the UFC’s early years was characterized by a small roster and single-night tournaments, naturally fostered frequent interactions among the same fighters~\cite{Gentry2011, Jennings2021MixedUFC}. This repetitive matchmaking amplified the network’s density, especially in the undirected network, which captures all interactions regardless of outcomes. The directed network, by contrast, reflects the competitive hierarchy, with slightly lower density due to its focus on win/loss relationships, limiting the total number of realized connections.

As the UFC transitioned into the Zuffa administration era, the density in both matchmaking-based networks steadily declined. By the window index of approximately 50, the undirected network had decreased to $2.5\times10^{-2}$, while the directed network stabilized at slightly lower values. This decline aligns with significant organizational reforms, including roster expansion and diversified matchmaking. The inclusion of new fighters diluted the proportion of realized connections, while Zuffa’s focus on globalizing the UFC introduced more diverse matchups, further reducing density (see Sec.~\ref{sec:introduction}). Fighters were no longer repeatedly paired with the same opponents, and rivalries became less frequent, reflecting a shift towards professionalized matchmaking aimed at competitive balance and narrative appeal. By the window index of 100, the density in both networks continued to decline, stabilizing around $0.5\times10^{-2}$ for the undirected network and slightly lower for the directed network. This stabilization marks the UFC’s transition to a decentralized and globally diverse structure, where only a small fraction of all potential matchups occur. Similar trends are observed in biological and seismic networks, where increased complexity leads to sparser connections \cite{Abe2004, Pasten2017, Hu2008}.

The winners and losers networks, in Fig.~\ref{fig:densityMethod2} reveal additional insights into the structural dynamics of victory and defeat. The undirected winners network begins with the highest density, starting at approximately $2.5\times10^{-1}$ and declining steadily to stabilize around $1.0\times10^{-1}$. This trend reflects the persistent connectivity of successful fighters, who often share opponents and form tightly-knit subgroups within the network. By contrast, the undirected losers network shows lower density, beginning at approximately $1.5\times10^{-1}$ and stabilizing around $0.5\times10^{-1}$. This disparity highlights the differing roles of winners and losers within the UFC’s competitive landscape, with victorious fighters maintaining a more central role in matchmaking strategies.

The directed winners and losers networks exhibit even lower densities, emphasizing the hierarchical nature of sequential victories and defeats. The directed winners network starts at approximately $1.0\times10^{-1}$ and stabilizes near $0.5\times10^{-1}$, capturing the dominance of specific fighters over time. In contrast, the directed losers network begins at around $0.5\times10^{-1}$ and stabilizes at much lower values, reflecting the transient and dispersed nature of defeat. These trends reveal how success is more likely to produce clusters, while defeats contribute to a broader, less cohesive structure.

When comparing the matchmaking-based networks to the winners and losers networks, it becomes evident that the UFC’s structural dynamics are shaped by contrasting narratives of success and defeat. The winners networks, particularly the undirected configuration, consistently exhibit higher densities, reflecting the organization’s emphasis on sustaining the visibility of successful fighters. In contrast, the losers networks show consistently lower densities, illustrating the decentralized and transient nature of defeat, as losing fighters are less likely to form persistent connections or remain central to the network. The directed networks for both winners and losers further highlight the hierarchical dynamics of sequential outcomes, where success leads to concentrated subgroups and defeat results in more distributed interactions.

%The historical evolution of these networks aligns closely with the UFC’s development. In its formative years, high densities in all networks reflect the organization’s small, centralized roster and repetitive matchmaking, creating a tightly-knit structure. Under Zuffa, the decline in density across all configurations underscores the UFC’s transition to a more professionalized and globally diversified sport. The winners networks, with their consistently higher densities, emphasize the UFC’s strategic focus on promoting dominant narratives and sustaining fan interest through successful fighters. The losers networks, by contrast, reflect the broader distribution of competitors and the transient nature of defeat within the UFC’s evolving ecosystem.

As the density values stabilize in later windows, the UFC demonstrates its ability to balance diversity with the promotion of key narratives. While lower density reflects a decentralized network with broader matchups, the persistence of higher densities in winners networks underscores the importance of dominant fighters and rivalries in sustaining audience engagement.

\begin{figure}[h!] \centering \includegraphics[width=\textwidth]{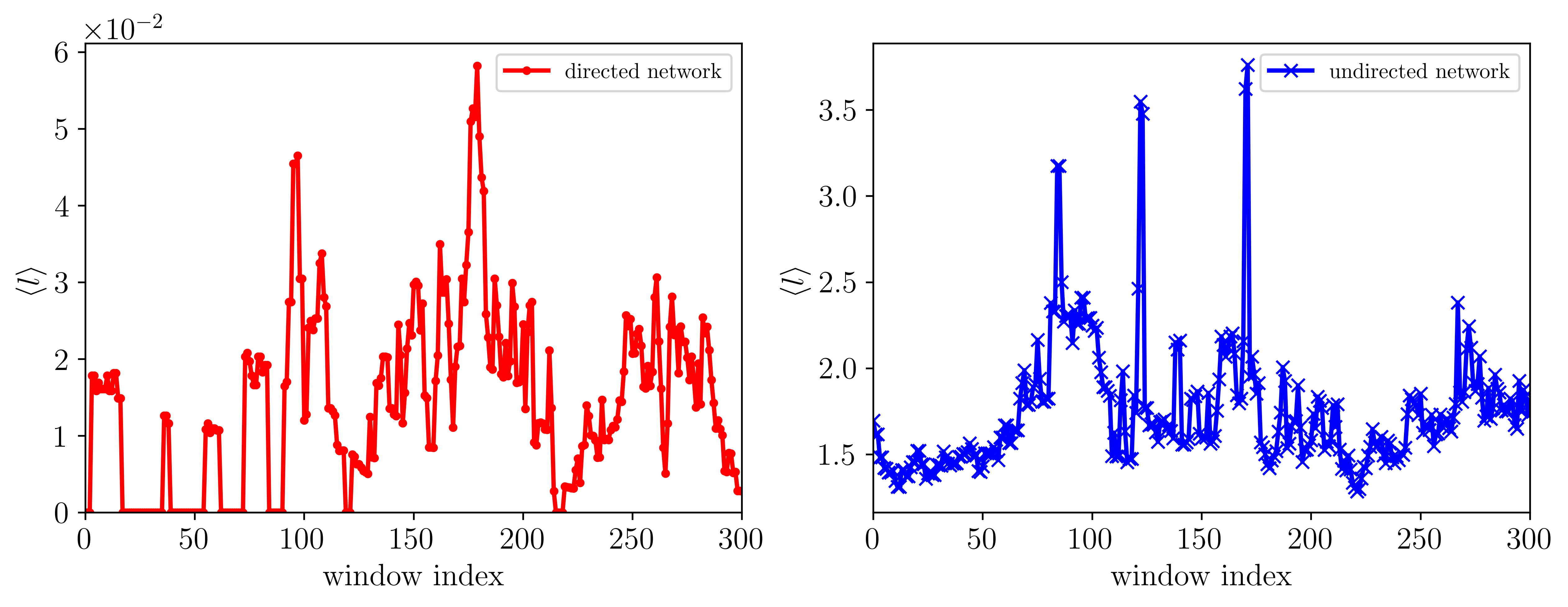} \caption{Average path length for the directed and undirected networks based on matchmaking, with directed values reduced by a factor of $10^{-2}$ due to averaging over disconnected components.} \label{fig:PathLenghtMethod1} \end{figure}

\begin{figure}[h!] \centering \includegraphics[width=0.7\textwidth]{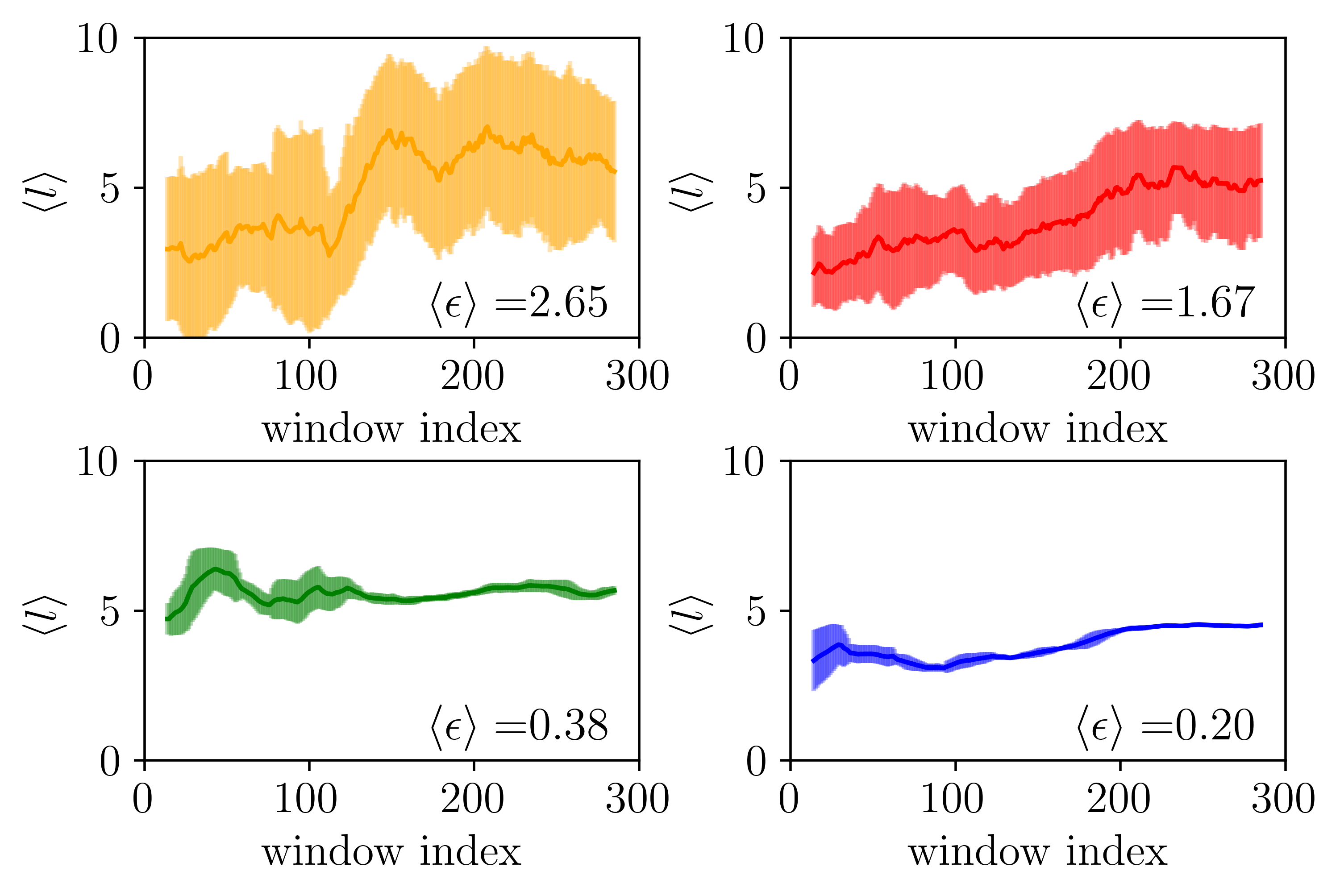} \caption{Average path length for winners and losers networks, with standard deviation values included due to data noise. (top left) losers directed network. (top right) winners directed network. (bottom left) losers undirected network. (bottom right) winners undirected network.} \label{fig:PathLengthWinnersLosers} \end{figure}

The evolution of the average path length $\langle l \rangle$ in Fig.~\ref{fig:PathLenghtMethod1} presents the trends for both directed and undirected networks based on matchmaking, while Fig.~\ref{fig:PathLengthWinnersLosers} illustrates the path length evolution for winners and losers networks. A critical observation in this analysis is the significant difference in magnitude between the directed and undirected networks in Fig.~\ref{fig:PathLenghtMethod1}, with the directed network consistently lower by a factor of $10^{-2}$ due to the presence of multiple disconnected components. This discrepancy arises from the necessity of computing path lengths individually for these disconnected components before averaging them. Despite this magnitude difference, both networks exhibit similar peak locations, indicating that the underlying structural changes affecting network connectivity are driven by common mechanisms.

Initially, the path length is relatively low in both networks, with the undirected network exhibiting values around 2.5, while the directed network remains substantially smaller due to averaging over isolated components. This early-stage low path length reflects the UFC’s formative years, where a smaller roster and frequent matchups among the same fighters created a tightly-knit network. Fighters were separated by only a few intermediaries, reinforcing dense connectivity~\cite{Newman2001, Barabasi1999}. The directed network’s lower values indicate the presence of disconnected clusters, where subgroups of fighters had no direct or indirect links to the broader network. These isolated components, often consisting of fighters who had only competed in one or two events before exiting the organization, contributed to the necessity of separate path length calculations.

As time progresses, the average path length increases, corresponding with the UFC’s roster expansion and matchmaking diversification. By the window index of 100, both networks exhibit distinct peaks, demonstrating that the structural evolution affects both broad connectivity and the competitive hierarchy. Despite the directed network’s lower values, its peaks align with those in the undirected network, suggesting that similar mechanisms—such as the introduction of weight classes, the influx of international fighters, and strategic matchmaking—drive these fluctuations~\cite{Gentry2011, Jennings2021MixedUFC}. The undirected network, reflecting general connectivity trends, follows a smooth upward trajectory, whereas the directed network exhibits pronounced variability due to the fragmentation caused by disconnected components.

The winners and losers networks is shown in Fig.~\ref{fig:PathLengthWinnersLosers}. It is important to note that due to heightened noise within this data, it has been averaged over a rolling window of six months, hence the error bars. In addition, the standard deviation $\langle \epsilon \rangle$ for each dataset has been written within each plot.

The directed winners network follows a similar trajectory to the matchmaking-based directed network but maintains slightly lower path lengths, indicating that successful fighters remain more interconnected over time. This suggests that elite competitors consistently face each other, forming a structured and dense competitive cluster. The directed losers network, in contrast, shows a higher path length, stabilizing around 6.5, with greater fluctuations in earlier windows. The greater path length in the losers network indicates that fighters who experience repeated losses are more likely to be part of a dispersed and transient connectivity structure, reinforcing their peripheral status in the UFC’s competitive hierarchy.

A deeper examination of the standard deviation values reveals crucial details about network stability and variability. The directed winners network maintains an average deviation of $\langle \epsilon\rangle = 1.67$, while the directed losers network shows a higher deviation of $\langle \epsilon\rangle = 2.65$. The larger variability in the losers network highlights the unpredictable nature of fighters who experience repeated losses. Unlike winners, who are consistently matched against other top competitors, losing fighters experience erratic matchmaking, often cycling through different opponents. This pattern suggests that defeat in the UFC leads to a more chaotic professional trajectory, with losing fighters more likely to be matched against newcomers or less-established competitors.

The undirected networks show a more stable structure, with significantly lower standard deviations. The undirected winners network has $\langle \epsilon\rangle=0.20$, while the undirected losers network maintains $\langle \epsilon\rangle=0.38$. The reduced variability in these networks suggests that the broader connectivity patterns of fighters—when not accounting for match outcomes—remain more stable over time. This reinforces the notion that while victories and losses influence connectivity within the directed framework, the overall matchmaking process remains structured, ensuring that fighters are consistently connected within the UFC ecosystem.

The distinction in standard deviations between winners and losers networks emphasizes the role of competitive success in shaping stable network trajectories. Winning fighters tend to remain within well-defined clusters, facing similarly ranked opponents, which leads to lower path length variability. Losing fighters, on the other hand, experience greater fluctuations in path length due to their transient position within the matchmaking structure. This aligns with the UFC’s promotional strategies, where successful fighters are kept within high-profile competitive clusters, while less successful fighters are often rotated through different levels of competition before being released from the organization.

%The historical implications of these findings align with the UFC’s transition from a tightly-knit organization to a global sport with a diverse matchmaking structure. The increasing path length, combined with declining density and clustering, suggests that while the UFC has expanded, it has also become less cohesive, with fighters experiencing fewer direct interactions. The directed network’s lower values highlight the fragmentation caused by disconnected components, reinforcing the hierarchical nature of competition where dominant fighters maintain strong connectivity while others become increasingly isolated.
From a strategic perspective, these trends indicate that the UFC tends to balance connectivity and diversity in matchmaking. While increasing path length reduces the likelihood of frequent rematches, it fosters a broader range of matchups, sustaining audience engagement. The stability observed in the winners network suggests that dominant fighters remain central to the organization’s strategies, ensuring that top competitors continue to generate high-profile bouts. The higher variability in the losers network reflects the UFC’s approach to cycling less-successful fighters through different competitive levels, maintaining a fluid and dynamic roster structure.
%-------------------------------------------

\begin{figure}[h!]
    \centering
    \includegraphics[width=\linewidth]{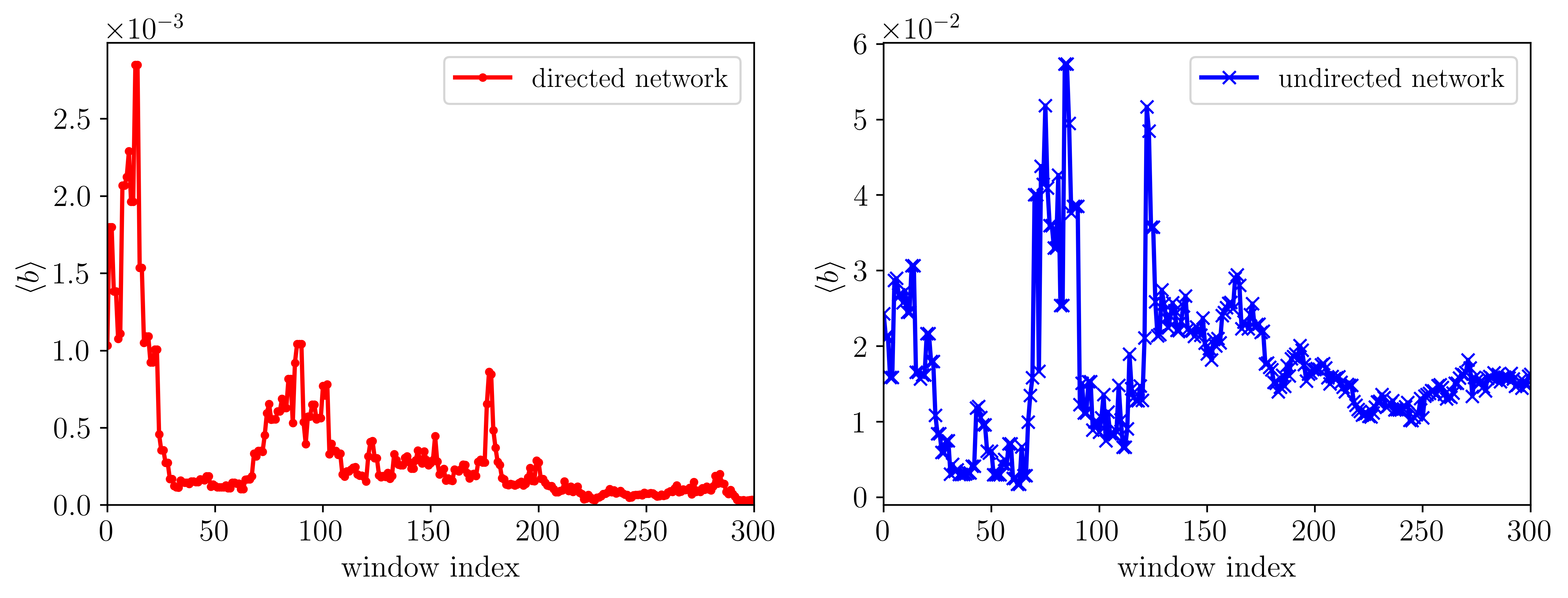}
    \caption{Betweenness centrality for the directed and undirected networks based on matchmaking.}
    \label{fig:betweenessMethod1}
\end{figure}

\begin{figure}[h!]
    \centering
    \includegraphics[width=0.7\linewidth]{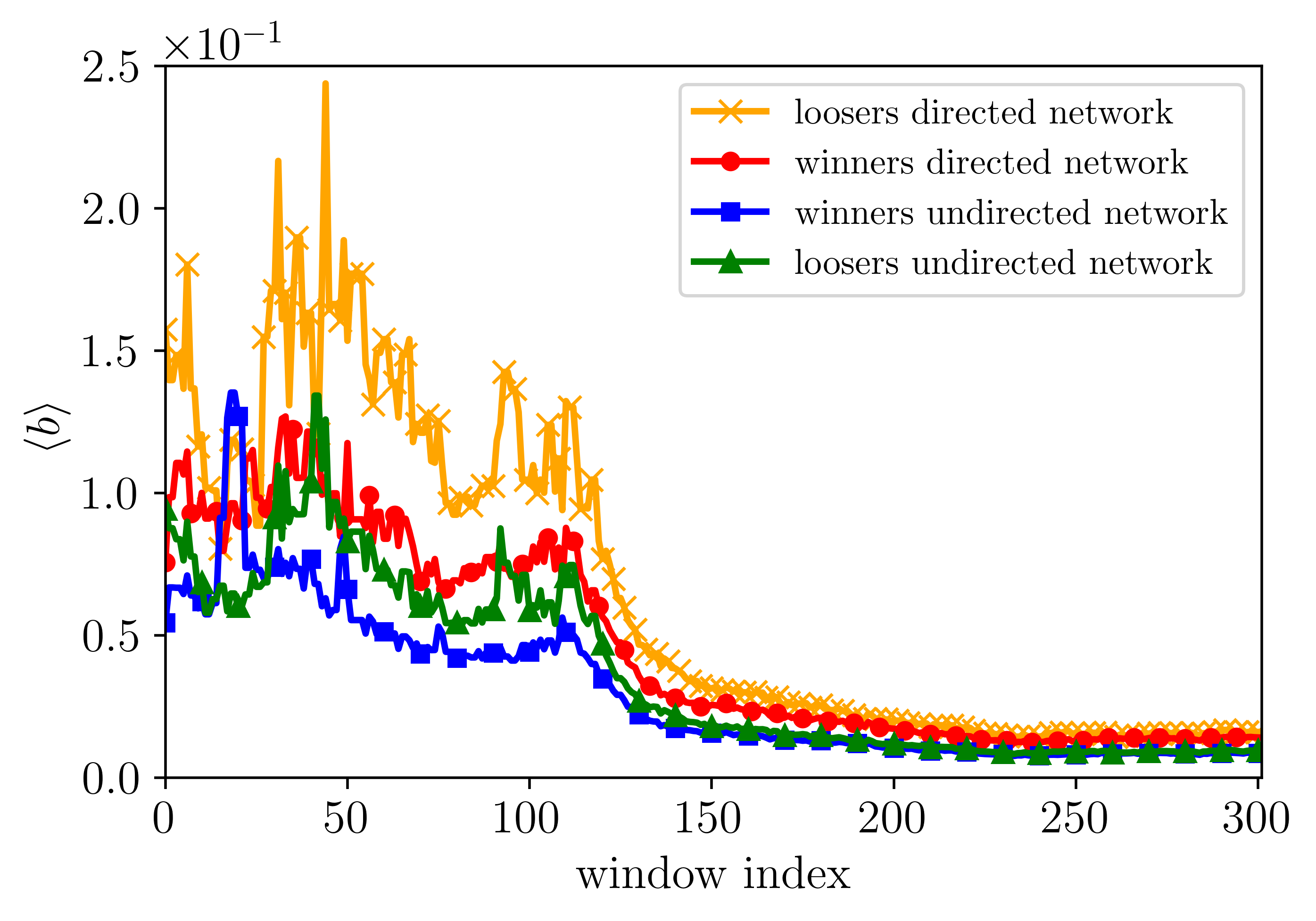}
    \caption{Betweenness centrality for the directed and undirected networks based on winners/losers networks.}
    \label{fig:betweenessMethod2}
\end{figure}

The evolution of betweenness centrality $\langle b \rangle$ in the UFC fighter network provides insights into how fighters serve as intermediaries connecting different subgroups within the network. Betweenness centrality is a crucial metric for identifying fighters who act as bridges between different competitive clusters, influencing the structural organization of the sport. Fig.~\ref{fig:betweenessMethod1} presents the results for the directed and undirected networks based on matchmaking, while Fig.~\ref{fig:betweenessMethod2} displays the winners and losers networks.

The directed network exhibits values that are consistently lower by a factor of $10^{-1}$ compared to the undirected network. This discrepancy arises from the presence of multiple disconnected components in the directed network, requiring individua calculations for each component before averaging. Despite this numerical difference, both networks exhibit similar behaviors in the locations of their peaks, indicating that key structural transitions in the UFC’s competitive landscape are captured in both network configurations. The undirected network, by aggregating all interactions regardless of outcome, provides a broader perspective on connectivity, while the directed network highlights the hierarchical nature of victories and losses, offering a more nuanced view of dominance and structural importance.

Initially, betweenness centrality is relatively low, with the undirected network starting around $2\times10^{-2}$. This suggests that in the early years of the UFC, the competitive structure was more evenly distributed, without a small subset of fighters disproportionately influencing connectivity. During this period, the UFC’s tournament-style events, with frequent rematches and a smaller roster, ensured that no single fighter played an overwhelmingly central role in connecting different parts of the network. As the UFC expanded, betweenness centrality increased, peaking near $1.7\times10^{-1}$ around the window index of 100. This peak reflects the emergence of dominant fighters who connected disparate groups of competitors, serving as key intermediaries in the network. This period corresponds with the rise of superstars who engaged in high-profile matches across multiple divisions, thereby shaping the structural cohesion of the UFC.

After reaching this peak, betweenness centrality steadily declines, stabilizing at approximately $0.2\times10^{-1}$ in later windows. This decline suggests a transition toward a more decentralized structure, where influence is distributed across a larger number of fighters rather than concentrated among a few key figures. This phase coincides with the expansion of the UFC’s roster, the introduction of new weight classes, and a diversification of matchmaking strategies, all of which contributed to the reduction of central dominance in the fighter network.

The winners and losers networks in Fig. \ref{fig:betweenessMethod2} provide additional insight into the hierarchical dynamics of UFC competition. The directed winners network exhibits a consistently moderate level of betweenness centrality, indicating that successful fighters play an essential role in bridging different competitive groups. These fighters often maintain their structural importance across multiple matchmaking cycles, ensuring that the competitive hierarchy remains stable. In contrast, the directed losers network initially exhibits higher betweenness centrality but declines rapidly. This suggests that losing fighters initially serve as structural intermediaries, likely due to their frequent participation in multiple bouts, but their long-term influence diminishes as they are gradually phased out of high-profile matchups.

The undirected winners and losers networks show more stable trends, with winners maintaining higher centrality than losers. This indicates that victorious fighters consistently play more significant roles in shaping the network structure, whereas losing fighters are less likely to remain pivotal over time. The fact that the undirected winners network retains a relatively stable centrality value highlights the UFC’s tendency to promote and match successful fighters in ways that reinforce their connectivity within the organization.

The peak in betweenness centrality coinciding with an increase in previous metrics suggests that this period was characterized by both greater connectivity and a concentration of influence among specific fighters. The subsequent decline in betweenness, alongside decreasing clustering coefficients and density, marks the UFC’s transition to a more distributed and less centralized structure. This reflects a shift from a reliance on a few dominant figures to a more dynamic competitive environment where matchmaking strategies distribute opportunities more evenly across the roster.

%These results provide critical insights into the UFC’s structural evolution. During its early years, the tightly connected network facilitated the rapid emergence of dominant figures who played crucial roles in matchmaking narratives. However, as the organization expanded and professionalized, it transitioned into a more balanced structure, where no single fighter or group maintained long-term central dominance. This decentralization likely contributed to the UFC’s ability to sustain viewer engagement over time, as the focus shifted from individual stars to a broader pool of talent. While this diversification fosters competitive fairness and allows new fighters to emerge, it also necessitates careful matchmaking strategies to maintain interest, as highly centralized networks with clear dominant figures have historically driven pay-per-view sales and fan engagement.

\begin{figure}
    \centering
    \includegraphics[width=0.7\linewidth]{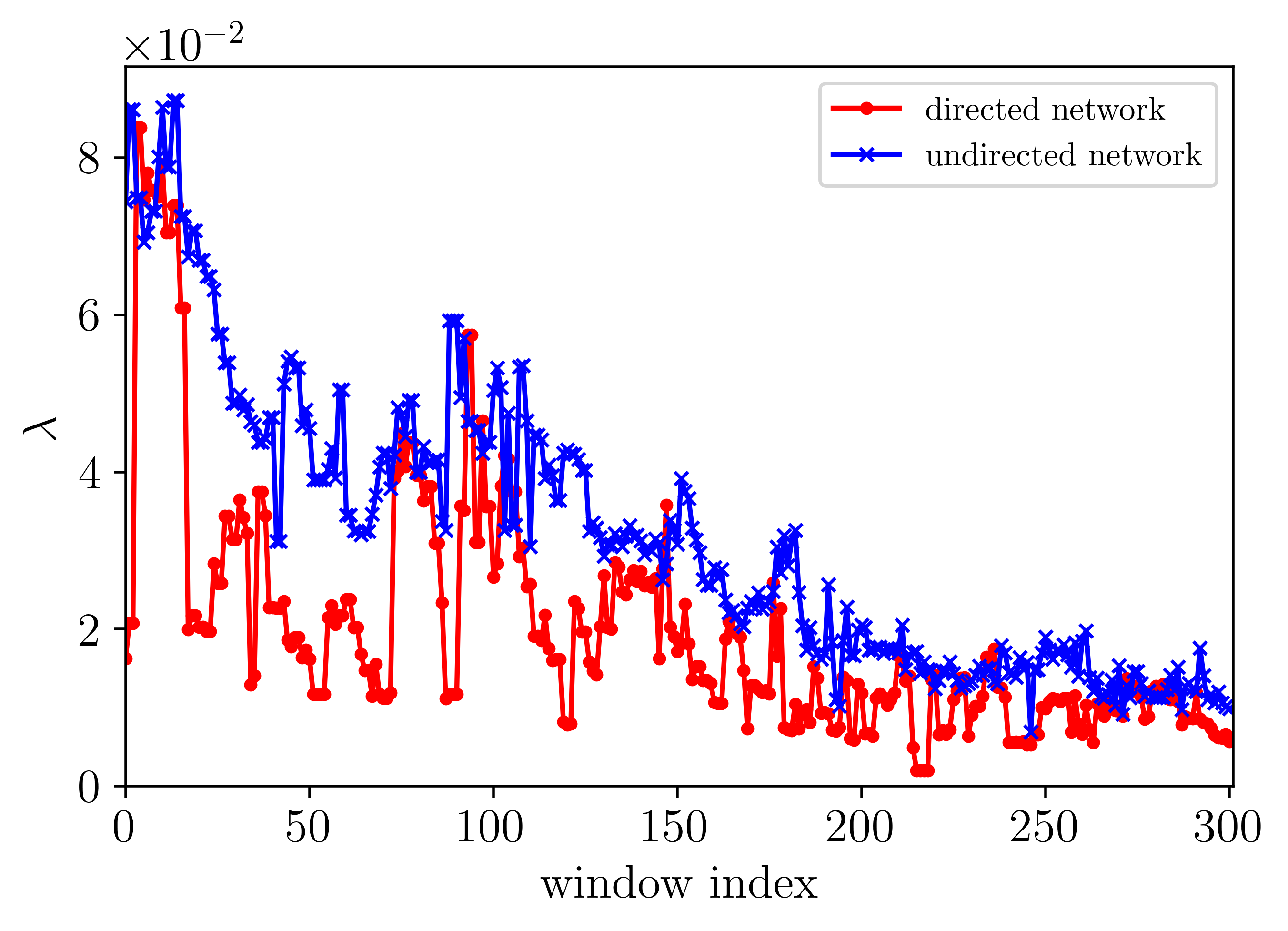}
    \caption{Eigenvector centrality for the directed and undirected networks based on matchmaking.}
    \label{fig:eigenvectorMethod1}
\end{figure}

\begin{figure}
    \centering
    \includegraphics[width=0.7\linewidth]{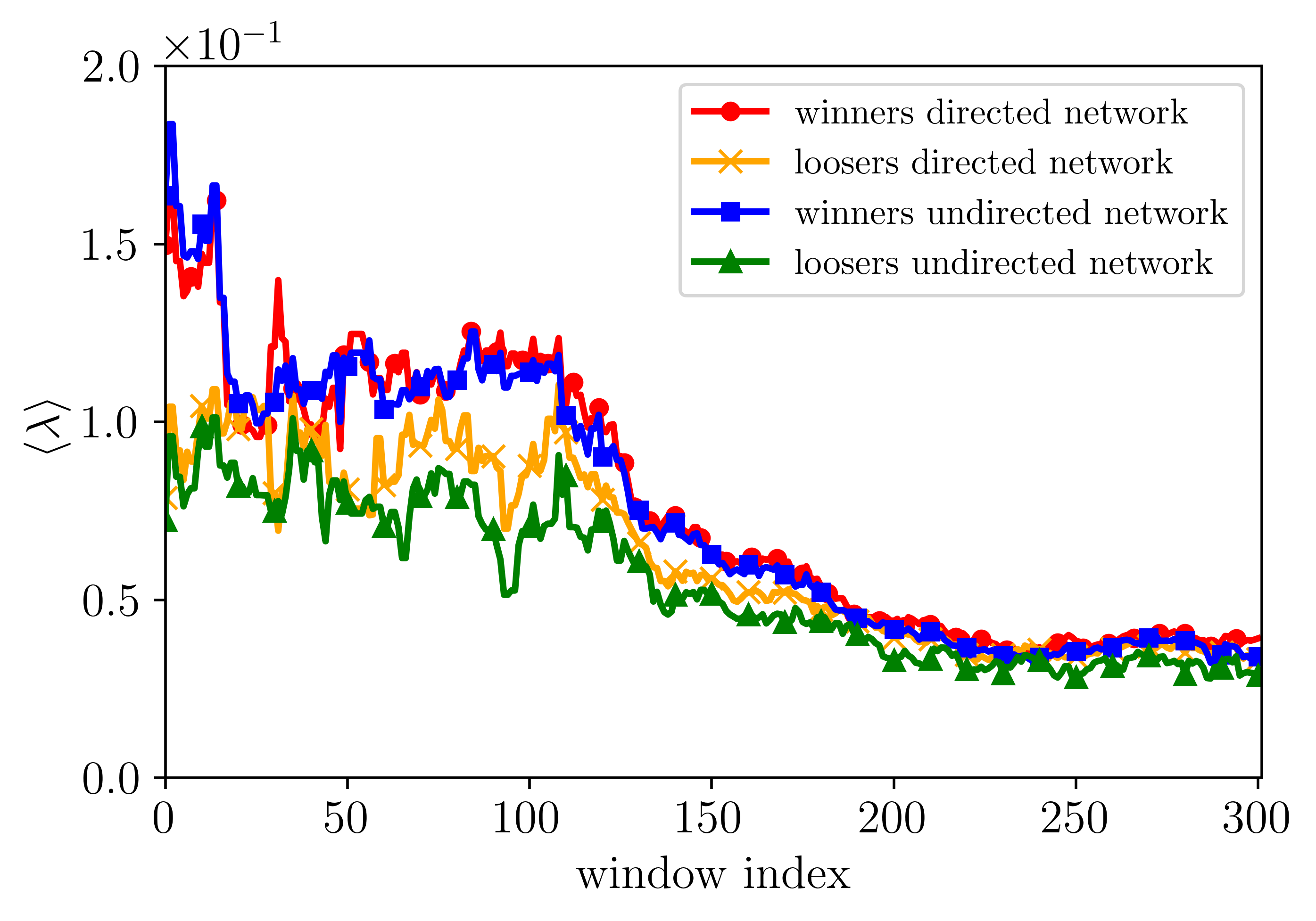}
    \caption{Eigenvector centrality for the directed and undirected networks based on winners/losers networks.}
    \label{fig:eigenvectorMethod2}
\end{figure}

The evolution of the average eigenvector centrality $\langle \lambda \rangle$ in the UFC fighter network reveals important insights into the shifting dynamics of influence within the competitive landscape. In this context, eigenvector centrality measures the extent to which a fighter is connected to other highly connected fighters, capturing a more nuanced view of influence beyond mere participation in fights. Fig.~\ref{fig:eigenvectorMethod1} illustrates the trends for the directed and undirected networks based on matchmaking, while Fig.~\ref{fig:eigenvectorMethod2} presents the results for the winners and losers networks.

Initially, the eigenvector centrality is moderately high, starting at approximately $1.0\times10^{-1}$, suggesting that even in the early stages of the UFC, certain fighters held significant influence by virtue of their connections to other well-connected competitors. This is consistent with the tightly clustered nature of the early UFC, where a smaller roster and frequent rematches reinforced connectivity among fighters. During this phase, high-profile fighters such as Royce Gracie dominated the network, maintaining strong ties with multiple competitors who themselves had extensive fight histories, thereby amplifying their centrality~\cite{Gentry2011, Jennings2021MixedUFC}.

As the window index progresses, eigenvector centrality increases, peaking around $2.0\times10^{-1}$ near the window index of 50. This peak reflects a period of consolidation in the network, where certain fighters rose to prominence not only by participating in a high volume of fights but by engaging with other well-connected fighters. The directed network shows consistently lower values compared to the undirected network, reinforcing the idea that directional fight outcomes create a more fragmented structure, where losses diminish a fighter’s centrality relative to those with sustained winning streaks. Despite these differences in scale, the directed and undirected networks exhibit similar trends, particularly in the locations of peaks and fluctuations, suggesting that structural shifts in the UFC's matchmaking strategy affected both representations.

Beyond the window index of 50, the eigenvector centrality begins to decline, though it remains punctuated by fluctuations, reflecting the dynamic nature of influence within the UFC. The peaks observed in both networks correspond to periods where specific fighters temporarily gained prominence, likely due to championship reigns, rivalries, or an increase in activity within certain weight classes. These fluctuations suggest that while some fighters achieved temporary dominance, their influence was frequently challenged by emerging competitors, leading to a continuous reshaping of the network.

After the window index of 100, eigenvector centrality exhibits a more sustained downward trend, stabilizing at approximately $0.2\times10^{-1}$ in later windows. This trend suggests a transition toward a more decentralized structure, where no single fighter or group of fighters maintains long-term dominance. The stabilization at lower values indicates that while influence remains a factor, it is more evenly distributed across a broader range of fighters, reducing the structural reliance on a few highly connected individuals. This shift aligns with other network metrics, such as decreasing betweenness centrality and increasing path length, which together suggest that the UFC evolved from a network dominated by a handful of elite fighters to one where influence is more widely dispersed.

The winners network, in Fig.~\ref{fig:eigenvectorMethod2}, consistently exhibits higher eigenvector centrality than the losers network, highlighting the structural advantage of victorious fighters in maintaining connectivity within the UFC’s competitive framework. In the directed winners network, centrality remains relatively stable, indicating that successful fighters maintain their influence across multiple matchmaking cycles. By contrast, the directed losers network exhibits a sharper decline in centrality over time, reflecting the transient nature of losing fighters' influence. Fighters who accumulate losses are less likely to remain central within the network, as their frequency of matchups against other influential fighters diminishes.

The undirected winners and losers networks show more moderate differences, but the same trend holds: winning fighters tend to maintain stronger network positions, while losing fighters gradually fade from prominence. The convergence of centrality values across all networks in later windows further supports the idea that the UFC has shifted toward a more balanced structure, where influence is more evenly distributed. This reflects an organizational strategy that promotes a broad array of fighters rather than concentrating marketability and matchmaking around a small elite group.

These findings suggest significant implications for the UFC’s strategic direction. The peak in eigenvector centrality during the middle period corresponds to a time when certain fighters were heavily promoted, driving pay-per-view sales and audience engagement through intense rivalries and championship narratives. However, as the centrality values declined and stabilized, the UFC likely adapted its promotional strategies to focus on a more diverse pool of athletes. This decentralization ensures sustainability by preventing over-reliance on a few marquee names while maintaining competitive intrigue through a dynamic and evolving network of contenders.

All these metrics collectively reflect the UFC’s evolution from a tightly organization to a globally diversified sport. During its early years, high density, clustering, and low path lengths highlight the compact and repetitive structure of the roster, where a few dominant fighters shaped the competitive hierarchy. As the UFC expanded under Zuffa, the network became more decentralized, with declining density and clustering coefficients reflecting the organization’s efforts to diversify competition. The peaks observed in betweenness and eigenvector centrality during the middle period correspond to the rise of iconic fighters who bridged different competitive clusters, driving audience engagement through high-profile narratives. Later, the stabilization of these metrics reflects the UFC’s maturity as a global sport. The increasing path lengths and declining centrality metrics highlight the organization’s transition to a balanced structure, where influence is distributed among a broader pool of fighters. This diversification reflects the UFC’s ability to sustain long-term interest by promoting a dynamic and evolving ecosystem of competitors.
%--------------------------------------------------
\subsection{Metrics correlations: the PPV sales and the Google search interest}

The correlation between network metrics and two key indicators of public engagement—PPV sales and Google search interest \cite{Tapology2024,Google2024}—provides valuable insights into how competitive structures influence the UFC’s commercial appeal and fan interest. While PPV sales represent a direct financial measure of event success, Google search interest reflects a broader scope of public curiosity, encompassing attention to upcoming fights, athlete controversies, and historical rivalries. Given the structural similarities among these time series, some degree of overlap in their trends is expected.

To quantify these relationships, we calculate the \textit{Pearson correlation coefficient} between the temporal evolution of each network metric and the corresponding PPV and Google Trends time series, a method widely employed in numerous fields of research—ranging from biological networks to sociological data—to assess linear associations between different types of measurements \cite{Freedman2005, Schober2018}. This approach allows us to pinpoint patterns of commercial and public engagement that evolve in tandem with changes in the underlying fighter network structure.

\begin{figure}[h!] \centering \includegraphics[width = 0.7\textwidth]{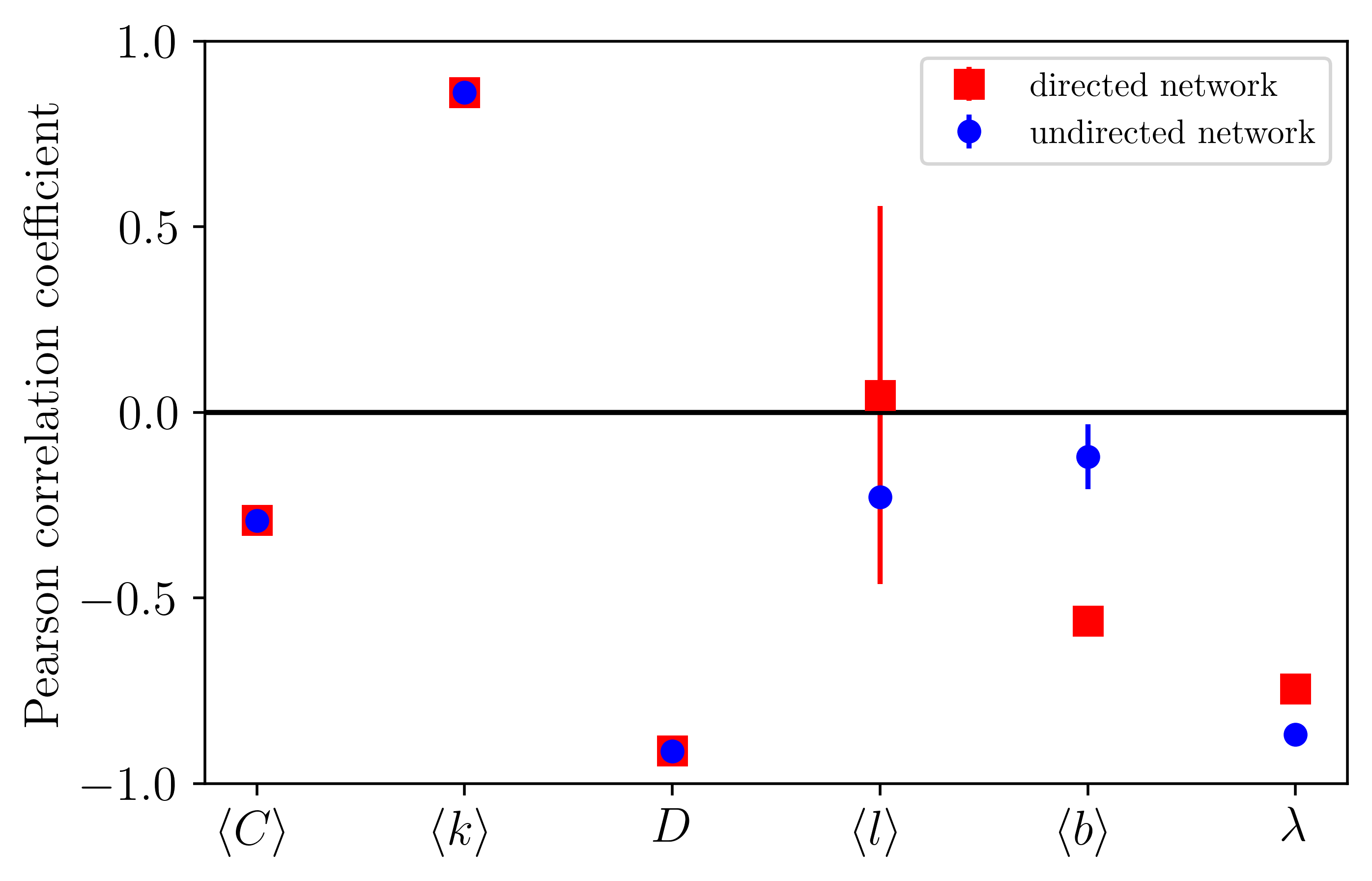} \caption{Pearson correlation coefficient between network metrics and PPV sales.} \label{fig:PPVcorr} \end{figure}

Fig.~\ref{fig:PPVcorr} illustrates the relationship between network evolution and PPV sales. The average degree $\langle k \rangle$ exhibits the strongest positive correlation, suggesting that when the network becomes more interconnected—i.e., when fighters engage with a wider pool of opponents—PPV sales increase. This supports the notion that highly dynamic matchmaking generates compelling narratives that enhance marketability. Conversely, network density $D$ shows a strong negative correlation, indicating that when the UFC network is highly interconnected, with fighters frequently facing common opponents, it diminishes the uniqueness of matchups, ultimately lowering PPV sales. This aligns with existing sports economics literature, which suggests that scarcity and novelty in rivalries drive audience interest~\cite{Duch2010}.

The clustering coefficient $\langle C \rangle$ follows a similar trend, showing a negative correlation with PPV sales. Highly clustered sub-networks—where fighters share many common opponents—may reduce the perceived novelty of fights, reinforcing the UFC’s strategic approach of limiting excessive rematches unless they involve highly anticipated rivalries. Meanwhile, the average path length $\langle l \rangle$ for the directed network correlates weakly with PPV sales, while in the undirected network this correlation is negative, implying that greater network dispersion, where fighters are separated by more intermediaries, sustains commercial interest. This likely reflects the UFC’s expansion strategy, where international matchmaking and cross-divisional fights add intrigue to events.

Both betweenness centrality $\langle b \rangle$ and eigenvector centrality $\lambda$ display weak to moderate negative correlations with PPV sales. This indicates that fighters acting as structural bridges between subgroups or those with high influence within the network do not necessarily drive commercial success. Instead, high-profile matchups, rather than network centrality, appear to be the key determinant of financial performance. This supports prior findings in complex network analyses of sports rankings, where structural influence does not always correlate with popularity or success metrics~\cite{Radicci2011}.

\begin{figure}[h!] \centering \includegraphics[width = 0.7\textwidth]{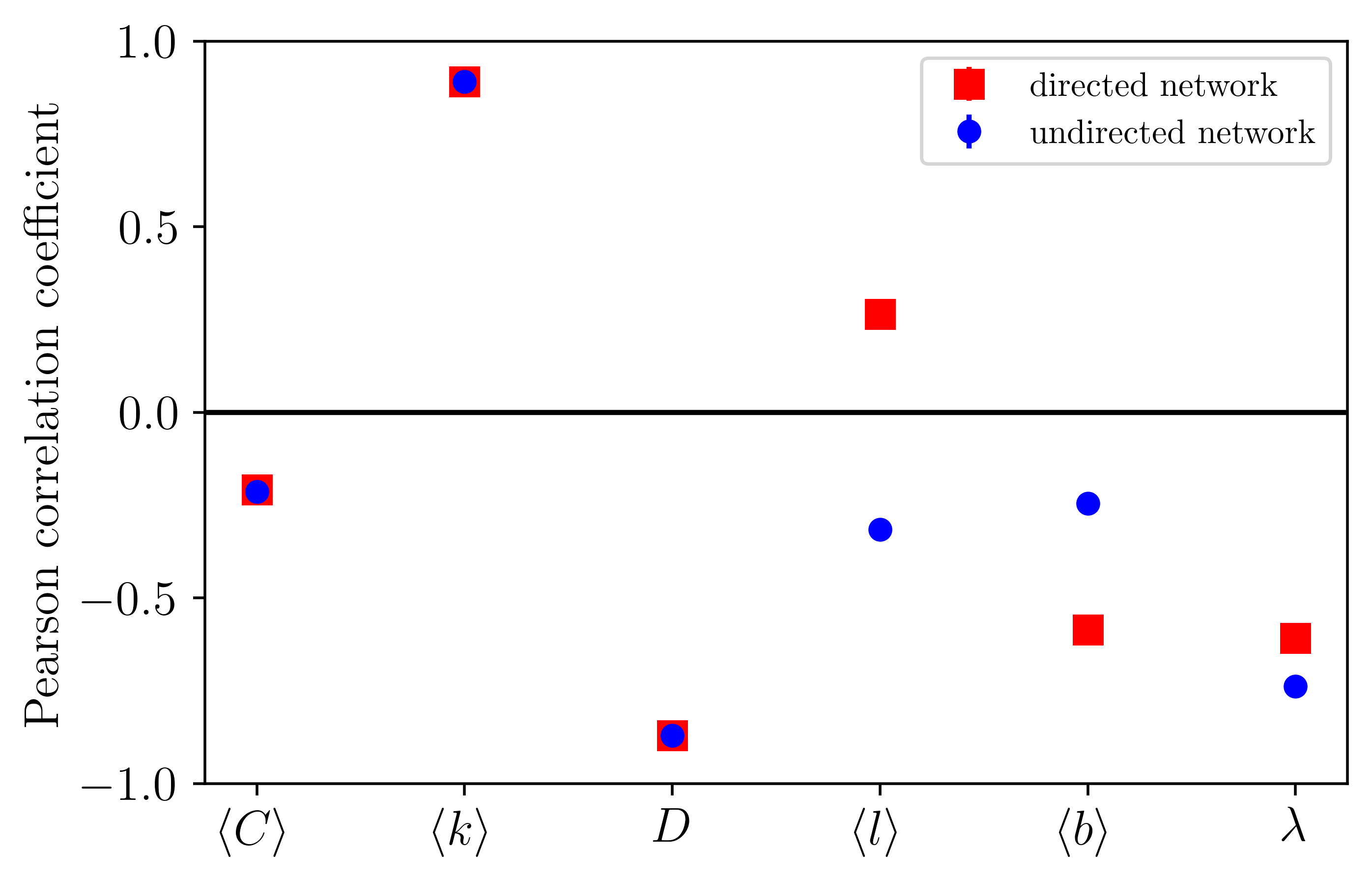} \caption{Pearson correlation coefficient between network metrics and Google search interest.} \label{fig:GoogleCorr} \end{figure}

Fig.~\ref{fig:GoogleCorr} presents the correlation between network metrics and Google search interest for the term \textit{"UFC"}. The general trends align with those seen in PPV sales, but with key distinctions. While average degree $\langle k \rangle$ retains a positive correlation, it is slightly weaker, suggesting that while network connectivity fuels fan engagement, external factors such as media coverage, fighter personalities, and controversies also contribute to search trends.

As with PPV sales, network density $D$ exhibits again a strong negative correlation, reinforcing the idea that excessive interconnectedness reduces the uniqueness of matchups. However, the clustering coefficient’s correlation with Google searches is weaker than with PPV sales, indicating that while highly clustered groups may not be optimal for generating revenue, they do not necessarily deter casual fan engagement.

Average path length $\langle l \rangle$ displays a low correlation and opposite behaviors between the directed and undirected networks. This suggests that while competitive separation contributes to commercial intrigue, search behaviors may be influenced by more immediate factors, such as social media trends and promotional efforts. Betweenness centrality $\langle b \rangle$ exhibits a weak positive correlation, in contrast to its negative relationship with PPV sales, suggesting that structurally central fighters contribute more to sustained fan interest than direct event revenue. Lastly, eigenvector centrality $\lambda$ maintains a negative correlation, further reinforcing the notion that an over-centralized network dominated by a few figures does not necessarily drive broader engagement.

\begin{figure}[!h] \centering \includegraphics[width = 0.7\textwidth]{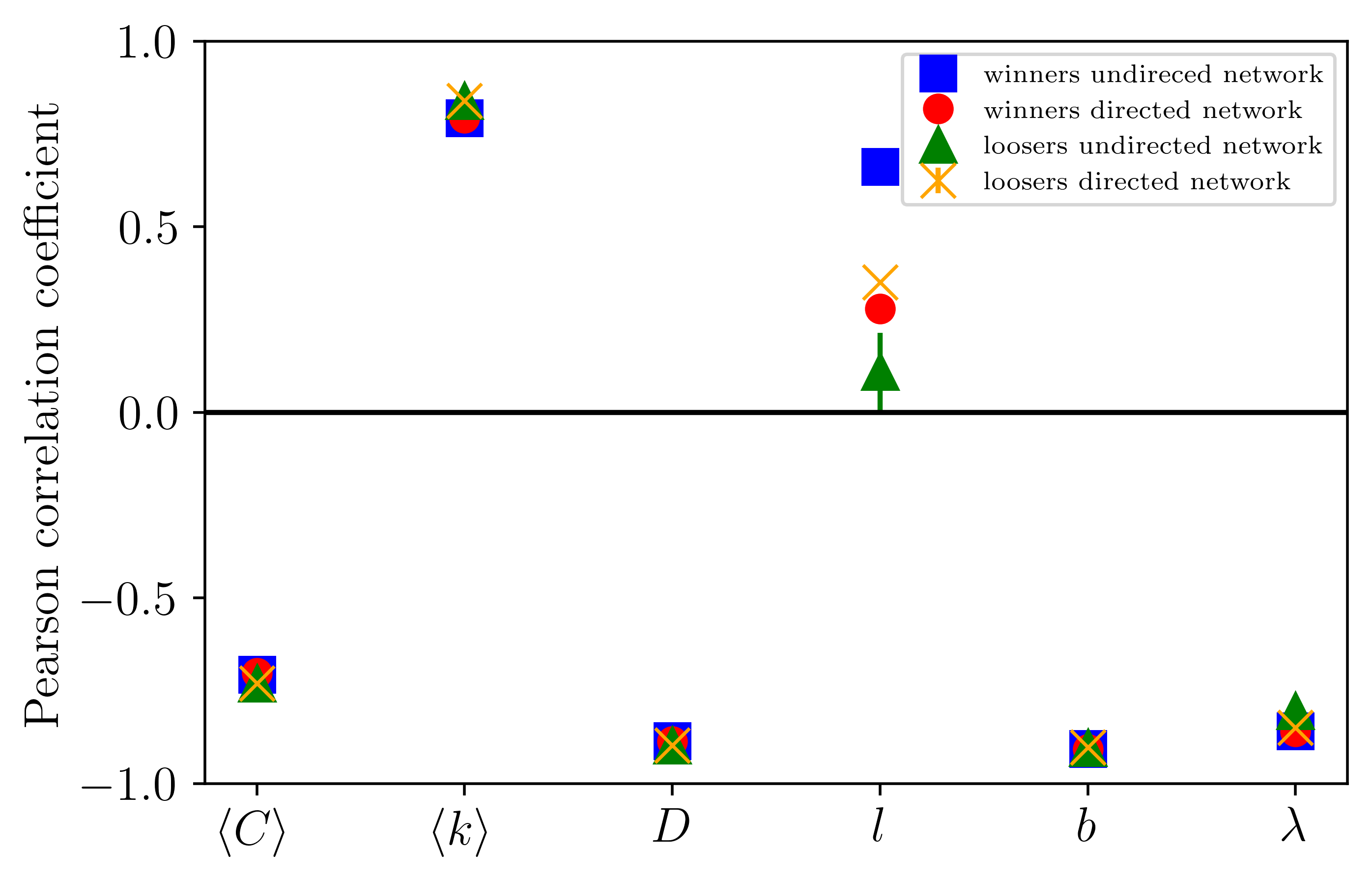} \caption{Pearson correlation coefficient between winners/losers network metrics and PPV sales.} \label{fig:PPVcorrMethod2} \end{figure}

Analyzing winners and losers networks in Fig.~\ref{fig:PPVcorrMethod2} provides additional insight. The average degree $\langle k \rangle$ exhibits a stronger positive correlation with PPV sales for the winners networks, emphasizing the commercial value of well-connected successful fighters. Conversely, the losers networks show a weaker but still positive correlation, implying that even the peripheral connectivity of less successful fighters contributes to the sport’s economic ecosystem.

Network density $D$ remains strongly negatively correlated across both winners and losers networks. Notably, the directed losers network exhibits the strongest negative correlation, reinforcing the idea that frequent matchups among lower-tier fighters do not drive PPV interest. Similarly, clustering coefficient $\langle C \rangle$ follows this trend, with a stronger effect in the winners networks, suggesting that repeated matchups among high-profile fighters reduce novelty and engagement.

A distinction emerges with average path length $\langle l \rangle$, which shows a moderate positive correlation. This indicates that greater separation between successful fighters before they eventually face each other enhances commercial anticipation. Meanwhile, both betweenness centrality $\langle b \rangle$ and eigenvector centrality $\lambda$ show strong negative correlations, particularly in the losers networks, reinforcing the idea that connectivity among frequently losing fighters does not contribute positively to PPV success.

\begin{figure}[!h] \centering \includegraphics[width = 0.7\textwidth]{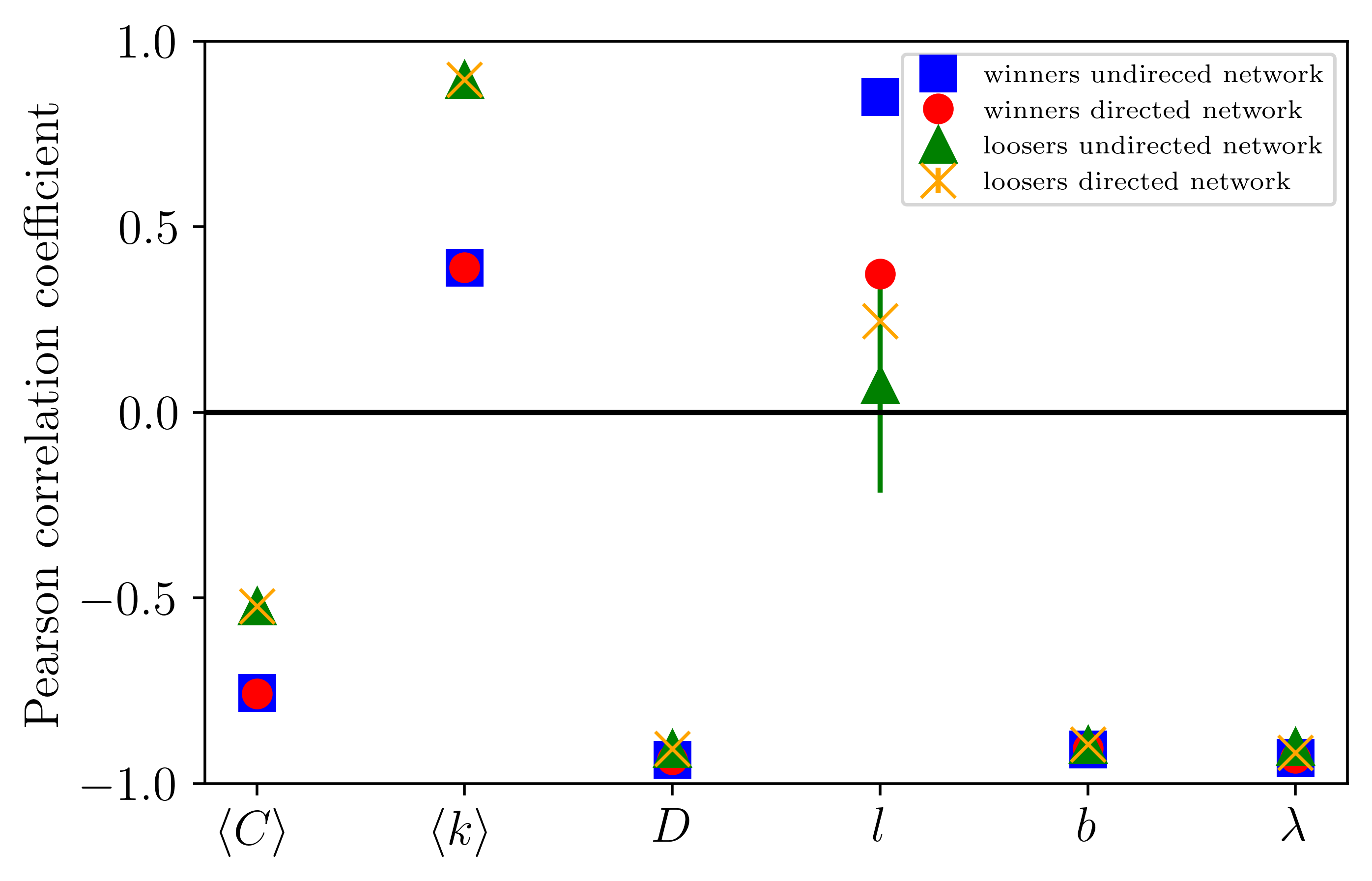} \caption{Pearson correlation coefficient between winners/losers network metrics and Google search interest.} \label{fig:GoogleCorrMethod2} \end{figure}

Fig.~\ref{fig:GoogleCorrMethod2} illustrates the correlation between network metrics and Google search interest for the winners and losers networks. As in the PPV sales analysis, the average degree $\langle k \rangle$ retains a strong positive correlation across all networks, with the winners networks exhibiting a more pronounced effect. This suggests that public interest in the UFC is sustained when dominant fighters remain actively engaged in a diverse set of matchups. The losers networks also show a positive correlation, though weaker, indicating that while losing fighters contribute to overall fan engagement, they do not sustain long-term attention as effectively as their more successful counterparts.

Network density $D$ remains negatively correlated across both winners and losers networks, reinforcing the idea that excessive interconnectedness dilutes the novelty of individual matchups. The strongest negative correlation is observed in the directed losers network, suggesting that repeated matchups among frequently losing fighters are less effective at generating sustained public interest. The clustering coefficient $\langle C \rangle$ follows a similar pattern, with negative correlations in both winners and losers networks. The slightly stronger effect observed in the losers networks suggests that when less-successful fighters compete within tightly-knit subgroups, it contributes less to overall audience engagement.

As the previous result, average path length $\langle l \rangle$, shows again a moderate positive correlation. Indicating that greater separation between successful fighters before they eventually face each other moderately enhances UFC interest.

Both betweenness centrality $\langle b \rangle$ and eigenvector centrality $\lambda$ exhibit consistently negative correlations across all networks, mirroring their relationship with PPV sales. This further supports the argument that commercial and public engagement are not necessarily driven by structurally central figures but rather by fighters with direct competitive success. The losers networks again show stronger negative correlations, reinforcing the notion that while structurally important fighters play a role in maintaining network cohesion, they are not the primary drivers of audience interest.

This results provides a comprehensive analysis of how UFC network structures correlate with public engagement, measured through PPV sales and Google search interest. The findings reveal that a highly interconnected network, reflected in average degree, consistently aligns with greater engagement, supporting the notion that active fighters competing across a diverse set of matchups sustain commercial appeal and fan curiosity. Meanwhile, network density and clustering coefficient negatively correlate with both PPV sales and search interest, reinforcing the importance of maintaining a balance between frequent interactions and the scarcity of high-profile matchups.

The winners vs. losers network analysis further highlights key distinctions. The stronger positive correlation between average degree and PPV sales in the winners networks suggests that connectivity among dominant fighters is a key driver of revenue generation. Conversely, the negative correlation between network density and engagement in the losers networks indicates that oversaturation of frequently losing fighters does not contribute positively to audience interest. Additionally, average path length emerges as an interesting factor, with greater separation between successful fighters correlating with higher PPV sales and sustained public interest in both winners and losers networks. This supports the UFC’s strategy of structuring long-term rivalries and building anticipation before marquee fights. While betweenness centrality and eigenvector centrality hold structural significance within the competitive framework, their negative correlations suggest that fighters with high network influence do not necessarily translate to financial or fan engagement success. Instead, the UFC’s promotional strategies appear to prioritize direct competitive success over network centrality.

It is important to emphasize that these findings represent an approximation of the complex dynamics between network structure and audience behavior. While clear trends emerge, further research is required to explore deeper causal mechanisms, including factors such as fighter-specific popularity, media exposure, and social media engagement. Future studies should integrate additional data sources and modeling approaches to refine our understanding of the UFC's matchmaking strategies and their broader impact on commercial success and public interest.

%-------------------------------------------------------
\subsection{Metric correlations: P4P, champions, and top-15 rankings}

\begin{figure}[!h]
    \centering
    \includegraphics[width = 0.9\textwidth]{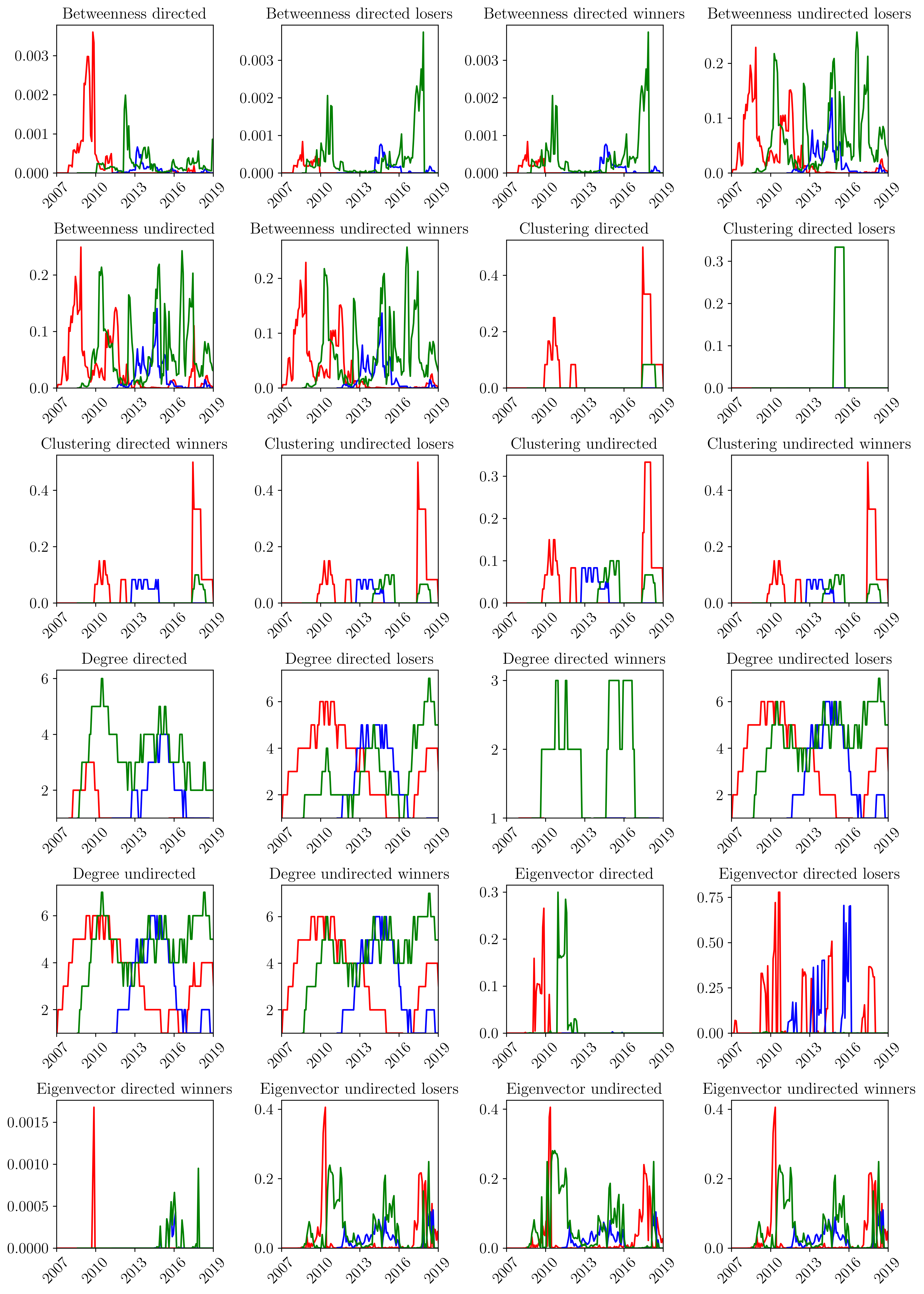}
    \caption{Evolution of the computed metrics per fighter. As an example, here we show in (red) Jon Jones, in (blue) Conor McGregor and in (green) Charles Oliveira.}
    \label{fig:metricsPerFighterExample}
\end{figure}

Some of the metrics introduced in Sec.~\ref{sec:methodMetrics} are computed at the node level. Specifically, in the following development, we evaluated the degree, clustering coefficient, betweenness centrality, and eigenvector centrality for each fighter across the constructed networks (for an example see Fig.~\ref{fig:metricsPerFighterExample}).

To explore whether network metrics align with competitive success in the UFC, we conducted an analysis comparing fighters ranked highest in key network metrics to those historically recognized as top competitors. This included pound-for-pound (P4P) rankings, divisional top 15 rankings, and UFC champions. By assessing these correlations, we aim to determine whether fighters with strong network positions—such as high connectivity, influence, or bridging roles—are more likely to be regarded as elite within the sport. All the used data can be accessed through Ref.~\cite{Tapology2024, Google2024, Warrier2021}.

For each network node metric (degree, clustering coefficient, betweenness centrality, and eigenvector centrality), we calculated the average value per fighter across all their UFC bouts to obtain a career-long metric score $M_j$ for a given fighter as follows,
\begin{equation}
    M_j=\frac{1}{N_{\text{fights}}}\sum_t m_j(t),
\end{equation}
where $m_j(t)$ is the value of metric $j$ at time $t$ and $N_{\text{fights}}$ is the total number of fights of the fighter. Fighters were then ranked based on these scores, prioritizing those with the highest values. These rankings were evaluated against three distinct success-based measures: the Pound-for-Pound (P4P) list, the UFC Champions, and the divisional Top-15 rankings. Notably, these ranking systems were officially introduced in early 2013; therefore, all comparisons were conducted using data from this period onward.

To quantify the overlap between fighters ranked highest in network metrics and those listed in competitive rankings, we used the Jaccard index, a similarity measure commonly applied in set theory and network science~\cite{Fletcher2018}. Given two sets $A$ and $B$, the Jaccard index is defined as:

\begin{equation}
    J(A,B) = \frac{|A\cap B|}{|A\cup B|}.
\end{equation}

In our context, $A$ represents the top fighters based on a given network metric, and $B$ represents the accumulated historical P4P, champions, or top-15 rankings. To ensure a fair comparison, if a ranking contained $X$ fighters, we selected the top $X$ fighters from each network metric ranking before computing the Jaccard index.

The results are shown in Table~\ref{tab:metrics}, and indicate that the overall highest Jaccard index values are observed in the top-15 rankings, particularly for degree and clustering in both directed and undirected networks. This suggests that fighters who maintain high connectivity—facing a diverse set of opponents over time—are more likely to be consistently ranked. The clustering coefficient in undirected networks exhibits the a high Jaccard index of 0.461, reinforcing the idea that fighters who belong to tightly-knit competitive subgroups tend to remain highly ranked, as their recurring matchups keep them relevant within their respective divisions.

\begin{table}[!h]
    \centering
    \begin{tabular}{lccc}
        \toprule
        Metric per fighter & Jaccard P4P & Jaccard ranked 0-15 & Jaccard champions\\
        \midrule
        $k$ und. & 0.089 & 0.443 & 0.043 \\
        $C$ und. & 0.065 & 0.461 & 0.029 \\
        $b$ und. & 0.077 & 0.322 & 0.059 \\
        $\lambda$ und.  & 0.065 & 0.322 & 0.043 \\
        $k$ und. winners & 0.089 & 0.438 & 0.043 \\
        $k$ und. losers & 0.089 & 0.438 & 0.043 \\
        $C$ und. winners & 0.077 & 0.455 & 0.029 \\
        $C$ und. losers & 0.077 & 0.455 & 0.029 \\
        $b$ und. winners & 0.101 & 0.325 & 0.075 \\
        $b$ und. losers & 0.101 & 0.325 & 0.075 \\
        $\lambda$ und. winners & 0.065 & 0.327 & 0.043 \\
        $\lambda$ und. losers & 0.065 & 0.327 & 0.043 \\
        $k$ dir. & 0.054 & 0.231 & 0.029 \\
        $C$ dir. & 0.101 & 0.330 & 0.059 \\
        $b$ dir. & 0.054 & 0.372 & 0.029 \\
        $\lambda$ dir. & 0.000 & 0.220 & 0.000 \\
        $k$ dir. winners & 0.032 & 0.162 & 0.014 \\
        $k$ dir. losers & 0.167 & 0.461 & 0.125 \\
        $C$ dir. winners & 0.089 & 0.383 & 0.059 \\
        $C$ dir. losers & 0.032 & 0.310 & 0.014 \\
        $b$ dir. winners & 0.021 & 0.352 & 0.000 \\
        $b$ dir. losers & 0.021 & 0.352 & 0.000 \\
        $\lambda$ dir. winners & 0.010 & 0.151 & 0.014 \\
        $\lambda$ dir. losers & 0.181 & 0.541 & 0.143 \\
        \bottomrule
    \end{tabular}
    \caption{Jaccard similarity index between P4P, top 15 and champions rankings and the undirected, directed, undirected winners, undirected losers, directed winners and directed losers per fighter network metrics.}
    \label{tab:metrics}
\end{table}

In contrast, the correlation between network metrics and P4P rankings is relatively weak. The highest Jaccard index for P4P fighters is 0.181 for eigenvector centrality in the directed losers' network, while most other values fall below 0.1. This suggests that the P4P ranking system, introduced in 2013, does not strongly align with network-based rankings, likely due to its subjective nature, which considers qualitative assessments of dominance, skill, and championship status rather than strictly structural connectivity~\cite{Robbins2017}. P4P rankings often favor long-reigning champions or fighters with notable cross-divisional achievements, factors that may not be adequately captured by network centrality measures.

Similarly, the analysis reveals that network-based rankings exhibit limited correlation with UFC champions. The highest Jaccard index for champions is 0.143 for directed eigenvector centrality in the losers’ network, while most values remain below 0.1. This finding suggests that championship success is not necessarily contingent on high connectivity within the network, likely due to matchmaking structures and promotional factors that influence the path to title contention. While network position may enhance visibility and opportunities, becoming a champion requires more than structural centrality, reinforcing previous research indicating that sports ranking systems often incorporate non-network variables such as skill assessment, historical performance, and promotional influence.

A particularly unexpected finding emerges in the analysis of the losers’ network, where degree and eigenvector centrality in the directed configuration show the strongest correlations with both ranking and championship status. With Jaccard indices of 0.167 for P4P, 0.541 for top 15 rankings, and 0.143 for champions, this suggests that maintaining network connectivity—even in a losing capacity—can sustain a fighter’s prominence within the UFC. This aligns with the promotion’s tendency to retain well-known fighters in high-profile fights despite their losses, as seen in the careers of figures like Donald Cerrone and Nate Diaz~\cite{Jennings2021MixedUFC}, whose popularity and continued presence in ranked fights persist despite inconsistent records. The importance of sustained exposure over strict win-loss records may explain why losing fighters with high connectivity retain ranking relevance, a phenomenon previously observed in network analyses of other professional sports leagues.

The broader implications of these results highlight the intricate interplay between network structure, competitive success, and ranking dynamics. Fighters who maintain high connectivity, particularly in terms of degree and clustering, are more likely to be ranked in the top 15 of their divisions, indicating that sustained activity and frequent matchups contribute to ranking stability. However, the weak correlations with P4P rankings suggest that broader assessments of fighter quality consider additional contextual factors beyond network positioning. The limited correlation with championship status further underscores the role of matchmaking and promotional influence in determining title contenders, factors that are not fully captured by network metrics.

It is crucial to recognize the limitations of this study. The Jaccard index measures direct set overlap and does not account for ranking order or the strength of competition. Additionally, network metrics alone do not capture stylistic advantages, promotional strategies, or external factors influencing fighter trajectories. Future research should explore weighted rankings, machine learning models, and historical progression analyses to develop a more comprehensive understanding of how network positioning influences success.

%-------------------------------------------------------

%-----------------------------

To examine the relationship between individual fighters' network characteristics and their public interest over time, we implemented a web-scraping algorithm to systematically extract Google Trends data for every fighter in the UFC database. This automated approach retrieved temporal search interest for each fighter, reflecting fluctuations in public attention across different periods. Following data collection, we computed the Pearson correlation coefficient between the Google search interest time series of each fighter and their degree, clustering coefficient, betweenness centrality, and eigenvector centrality per node. This allowed us to quantitatively assess how a fighter's network position within the competitive structure aligns with public engagement.

\begin{figure}[!h]
    \centering
    \includegraphics[width = 1\textwidth]{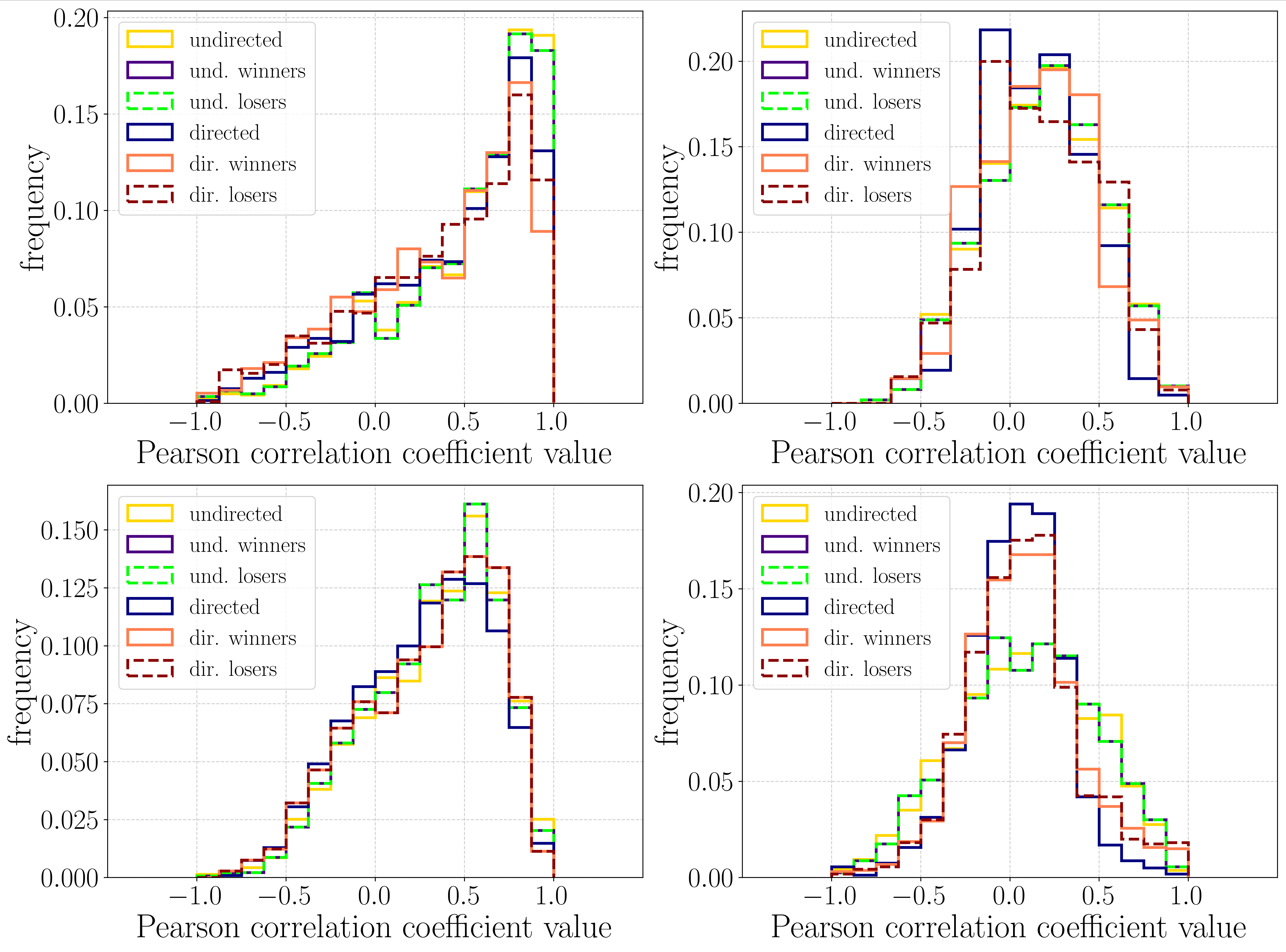}
    \caption{Histogram for the Pearson correlation coefficient between the evolution of the metrics per fighter and the Google interest per fighter.}
    \label{fig:histogramCorrelationTrends}
\end{figure}

The histograms in Fig.~\ref{fig:histogramCorrelationTrends} summarize the distribution of these correlation coefficients across different network configurations. The results are presented for directed and undirected networks, as well as the winners and losers networks, offering a comprehensive perspective on how different competitive trajectories influence public recognition.

The top-left histogram depicts the correlation between fighter degree and Google search interest. The observed distribution skews positively, suggesting that fighters with a higher degree tend to experience stronger public interest. This aligns with the notion that network connectivity directly influences popularity and media attention. Notably, the distributions are highly similar across directed, undirected, winners, and losers networks, indicating that degree remains a consistent predictor of public interest regardless of competitive trajectory.

In contrast, the top-right histogram represents the correlation between clustering coefficient and Google search interest. The clustering coefficient captures the local connectivity of a fighter, measuring the extent to which their opponents also compete against each other. The correlation distribution is centered around zero, with slight negative tendencies, particularly in the losers networks. This suggests that fighters embedded in tightly-knit sub-networks do not necessarily sustain high public engagement. The negative skew aligns with previous observations that high clustering often reduces novelty and limits audience excitement~\cite{Duch2010}. Fighters frequently competing within the same pool of opponents may not generate as much interest as those involved in diverse and unpredictable matchups.

The bottom-left histogram displays the distribution of Pearson correlation coefficients between betweenness centrality and Google search interest for individual UFC fighters. This metric is particularly relevant in combat sports, as it identifies fighters who facilitate connections between otherwise disconnected regions of the competition network~\cite{Radicci2011}.

The correlation distribution reveals a strong tendency toward positive values, indicating that fighters with higher betweenness centrality generally experience greater public interest over time. This suggests that fighters who frequently connect distinct groups of competitors—either by moving between weight divisions, facing diverse opponents, or participating in high-profile cross-divisional matchups—garner higher levels of Google search interest. These results reinforce the idea that fighters with diverse competitive trajectories and broad matchmaking histories tend to sustain public engagement more effectively than those whose careers are confined to a narrow set of repeated interactions.

Finally, the bottom-right histogram illustrates the distribution of Pearson correlation coefficients between eigenvector centrality and Google search interest for individual UFC fighters. Eigenvector centrality extends degree centrality by weighting a fighter’s connections based on the influence of their opponents, measuring how well-connected a fighter is within the broader competition network. High eigenvector centrality implies connections to other highly connected fighters, while lower values indicate more peripheral positions in the network~\cite{Radicci2011}.

The distribution of correlation values is centered around zero, indicating that eigenvector centrality has little systematic relationship with Google search interest. Unlike other network metrics such as degree or betweenness centrality, which exhibited clear tendencies toward positive correlations, eigenvector centrality does not appear to be a strong predictor of sustained public engagement. This suggests that being connected to highly influential fighters does not necessarily translate to higher search volume or sustained fan interest.

The absence of a strong correlation may stem from the nature of the UFC's promotional strategies and fan engagement mechanisms. While a fighter with high eigenvector centrality competes against well-connected opponents, their own individual visibility may remain secondary to the stars they are associated with. This contrasts with betweenness centrality, where a fighter’s ability to bridge different competitive subgroups appears to play a more significant role in determining long-term public interest.

The findings of this study highlight the nuanced relationship between network centrality metrics and public interest in UFC fighters. Degree and betweenness centrality exhibit the strongest positive correlations with Google search interest, suggesting that fighters who engage in a greater number of matchups and serve as competitive bridges between distinct subgroups tend to attract more sustained public attention. These results align with previous research in network analysis and sports analytics, where higher connectivity and diverse competitive exposure enhance an athlete’s visibility and long-term recognition.

In contrast, clustering coefficient and eigenvector centrality show little to no systematic relationship with search interest, implying that fighters embedded within tightly-knit subgroups or connected to highly influential opponents do not necessarily generate higher levels of engagement. This suggests that public interest in combat sports is driven more by direct exposure and career trajectory than by a fighter’s indirect influence within the competitive structure. The negative correlation observed for clustering coefficient further supports the idea that excessive redundancy in matchmaking—where fighters continuously compete against shared opponents—diminishes novelty and fan engagement.

These insights contribute to a broader understanding of how competitive structures shape public interest and commercial success in professional sports. They reinforce the idea that fighters who frequently transition between competitive clusters, take on diverse opponents, or participate in high-profile bouts sustain stronger long-term visibility than those whose careers remain confined to specific rivalries or divisions. The contrast between the predictive power of betweenness centrality and eigenvector centrality is particularly telling: while the former reflects a fighter’s role in connecting different competitive subgroups, the latter captures positional influence without necessarily driving independent public recognition. This distinction underscores the importance of active engagement over passive association with high-profile fighters.
%%%%%%%%%%%%%%%%%%%%%%%%%%%%%%%%%%%%%%%%%%
\section{Discussion}

The analysis of the UFC fighter network has revealed significant structural transformations that have taken place throughout the organization’s history. Initially, the networkS exhibited a highly interconnected structure, driven by a small roster and frequent rematches, particularly in the early years of single-night tournaments. Over time, as the UFC expanded and introduced new competitive policies, a marked decline in network density and clustering coefficient was observed. This decline aligns with the organization’s efforts to diversify matchmaking and reduce the dominance of a select group of fighters. Directed networks, which highlight the hierarchical nature of victories and losses, consistently demonstrated lower average degrees, emphasizing the concentration of dominance within a small subset of fighters.

Further insights emerged from the winners and losers networks, which revealed contrasting competitive trajectories. Fighters in the winners network maintained higher average degrees, clustering coefficients, and connectivity, underscoring their central role in the UFC’s matchmaking strategy. In contrast, the losers network exhibited lower degrees and less structural cohesion, highlighting the transient nature of defeat within the organization. These findings suggest that the UFC’s competitive ecosystem is designed to sustain the visibility of successful fighters while allowing for turnover among less successful competitors.

Metrics such as path length and betweenness centrality provided additional perspectives on the evolution of the UFC’s matchmaking strategy. The increase in average path length over time indicates a shift from a tightly interconnected structure to a more distributed and complex competitive landscape. Meanwhile, betweenness centrality peaked during the middle period of the organization’s evolution, corresponding to the rise of dominant fighters who acted as bridges between different competitive clusters. This phase was characterized by intense rivalries and high-profile matchups that generated significant audience interest. However, as the network matured, betweenness centrality declined, suggesting a transition toward a more balanced structure where influence is distributed across a larger number of fighters.

The correlation between network metrics and public engagement, measured through PPV sales and Google search interest, further substantiated the role of structural connectivity in shaping audience interest. The findings demonstrated that average degree positively correlates with both PPV sales and Google search interest, reinforcing the idea that broad competitive engagement sustains fan curiosity and marketability. Conversely, high network density and clustering coefficients exhibited negative correlations, indicating that excessive interconnectedness and repetitive matchups diminish commercial appeal. This aligns with previous research in sports economics, which emphasizes the importance of novelty in maintaining audience engagement.

A detailed examination of individual fighter metrics revealed that network positioning alone does not fully predict competitive success. While high connectivity and clustering correlate with divisional rankings, the correlation with Pound-for-Pound (P4P) rankings and championship status was relatively weak. This suggests that broader assessments of fighter quality incorporate additional factors such as technical skill, promotional appeal, and career trajectory beyond mere network structure. Interestingly, the directed losers network exhibited the strongest correlations with rankings and championship status, indicating that fighters who maintain competitive relevance—despite accumulating losses—often retain a degree of visibility within the organization.

Future research in this area should aim to refine the models of competitive success by incorporating additional variables such as fighter-specific popularity, media exposure, and social media engagement. Moreover, further exploration of matchmaking policies could provide deeper insights into the balance between competitive integrity and commercial interests. Expanding this network-based approach to other combat sports or even broader professional leagues could yield valuable comparative analyses, offering a comprehensive understanding of how organizational structures shape competitive trajectories. Ultimately, this study provides a foundational framework for analyzing the dynamics of sports networks, with implications for both strategic matchmaking and audience engagement.

\section*{Funding}
This research was funded by FONDECyT grant number 1240810 and FONDECyT grant number 1242013. 

\section*{Acknowledgments}
We would like to thank “Núcleo de Investigación No.7 UCN-VRIDT 076/2020, Núcleo de modelación y simulación científica (NMSC)” for the scientific support. Computational work was partially supported by the supercomputing infrastructures of the NLHPC (ECM-02).

%Bibliography
\bibliographystyle{unsrturl}   
\bibliography{references}

\begin{thebibliography}{10}

\bibitem{Robbins2017}
Thomas R.~Robbins and James E.~Zemanek, Jr.
\newblock {UFC pay-per-view buys and the value of the celebrity fighter}.
\newblock {\em Innovative Marketing}, 13(4):35--46, 12 2017.
\newblock \href {https://doi.org/10.21511/im.13(4).2017.04} {\path{doi:10.21511/im.13(4).2017.04}}.

\bibitem{Jennings2021MixedUFC}
L.A. Jennings.
\newblock {\em {Mixed Martial Arts: A History from Ancient Fighting Sports to the UFC}}.
\newblock Rowman {\&} Littlefield Publishers, 2021.

\bibitem{Snowden2010}
Jonathan Snowden.
\newblock {\em {Total Mma: Inside Ultimate Fighting}}.
\newblock 2010.

\bibitem{Kevlar2013}
{Kevlar}.
\newblock {Chris Weidman knock out Anderson Silva at UFC 162.}, 2013.
\newblock URL: \url{https://en.wikipedia.org/wiki/File:Chris_Weidman_knock_out_Anderson_Silva_at_UFC_162..jpg}.

\bibitem{Davie2014}
Art Davie.
\newblock {\em {Is This Legal?: The Inside Story of the First UFC from the Man Who Created It}}.
\newblock Ascend Books, 2014.

\bibitem{Gentry2011}
Clyde Gentry.
\newblock {\em {No Holds Barred: The Complete History of Mixed Martial Arts in America}}.
\newblock Triumph Books, 2011.

\bibitem{Schildhauer2013}
Angelo Schildhauer.
\newblock {\em {The Regulation of Mixed Martial Arts}}.
\newblock PhD thesis, Texas A{\&}M University, 2013.

\bibitem{Merced2016}
Michael~J. de~la Merced.
\newblock {U.F.C. Sells Itself for {\$}4 Billion}, 2016.
\newblock URL: \url{https://www.nytimes.com/2016/07/11/business/dealbook/ufc-sells-itself-for-4-billion.html}.

\bibitem{Newman2003}
M.~E.~J. Newman.
\newblock {The Structure and Function of Complex Networks}.
\newblock {\em SIAM Review}, 2003.
\newblock \href {https://doi.org/10.1137/S003614450342480} {\path{doi:10.1137/S003614450342480}}.

\bibitem{Barabasi2002b}
Réka Albert and Albert-László Barab{\'{a}}si.
\newblock {Statistical mechanics of complex networks}.
\newblock {\em Reviews of Modern Physics}, 74(1):47--97, 1 2002.
\newblock \href {https://doi.org/10.1103/RevModPhys.74.47} {\path{doi:10.1103/RevModPhys.74.47}}.

\bibitem{Radicci2011}
Filippo Radicchi.
\newblock {Who Is the Best Player Ever? A Complex Network Analysis of the History of Professional Tennis}.
\newblock {\em PLoS ONE}, 6(2):e17249, 2 2011.
\newblock \href {https://doi.org/10.1371/journal.pone.0017249} {\path{doi:10.1371/journal.pone.0017249}}.

\bibitem{Barabasi2002}
Réka Albert and Albert-László Barab{\'{a}}si.
\newblock {Statistical mechanics of complex networks}.
\newblock {\em Reviews of Modern Physics}, 74(1):47--97, 1 2002.
\newblock \href {https://doi.org/10.1103/RevModPhys.74.47} {\path{doi:10.1103/RevModPhys.74.47}}.

\bibitem{Stein2009}
Amelie Stein, Roland~A. Pache, Pau Bernad{\'{o}}, Miquel Pons, and Patrick Aloy.
\newblock {Dynamic interactions of proteins in complex networks: a more structured view}.
\newblock {\em The FEBS Journal}, 276(19):5390--5405, 10 2009.
\newblock \href {https://doi.org/10.1111/j.1742-4658.2009.07251.x} {\path{doi:10.1111/j.1742-4658.2009.07251.x}}.

\bibitem{Cherubini2015}
Christian Cherubini, Simonetta Filippi, Alessio Gizzi, and Alessandro Loppini.
\newblock {Role of topology in complex functional networks of beta cells}.
\newblock {\em Physical Review E}, 92(4):042702, 10 2015.
\newblock \href {https://doi.org/10.1103/PhysRevE.92.042702} {\path{doi:10.1103/PhysRevE.92.042702}}.

\bibitem{Albert2005b}
Réka Albert.
\newblock {Scale-free networks in cell biology}.
\newblock {\em Journal of Cell Science}, 118(21):4947--4957, 11 2005.
\newblock \href {https://doi.org/10.1242/jcs.02714} {\path{doi:10.1242/jcs.02714}}.

\bibitem{Costa2008}
Luciano da~F. Costa, Francisco~A. Rodrigues, and Alexandre~S. Cristino.
\newblock {Complex networks: the key to systems biology}.
\newblock {\em Genetics and Molecular Biology}, 31(3):591--601, 2008.
\newblock \href {https://doi.org/10.1590/S1415-47572008000400001} {\path{doi:10.1590/S1415-47572008000400001}}.

\bibitem{Braun2012}
Pascal Braun and Anne‐Claude Gingras.
\newblock {History of protein–protein interactions: From egg‐white to complex networks}.
\newblock {\em PROTEOMICS}, 12(10):1478--1498, 5 2012.
\newblock \href {https://doi.org/10.1002/pmic.201100563} {\path{doi:10.1002/pmic.201100563}}.

\bibitem{Spirin2003}
Victor Spirin and Leonid~A. Mirny.
\newblock {Protein complexes and functional modules in molecular networks}.
\newblock {\em Proceedings of the National Academy of Sciences}, 100(21):12123--12128, 10 2003.
\newblock \href {https://doi.org/10.1073/pnas.2032324100} {\path{doi:10.1073/pnas.2032324100}}.

\bibitem{Tucker2001}
C~Tucker.
\newblock {Towards an understanding of complex protein networks}.
\newblock {\em Trends in Cell Biology}, 11(3):102--106, 3 2001.
\newblock \href {https://doi.org/10.1016/S0962-8924(00)01902-4} {\path{doi:10.1016/S0962-8924(00)01902-4}}.

\bibitem{Araujo2007}
Robyn~P. Araujo, Lance~A. Liotta, and Emanuel~F. Petricoin.
\newblock {Proteins, drug targets and the mechanisms they control: the simple truth about complex networks}.
\newblock {\em Nature Reviews Drug Discovery}, 6(11):871--880, 11 2007.
\newblock \href {https://doi.org/10.1038/nrd2381} {\path{doi:10.1038/nrd2381}}.

\bibitem{Pasten2018}
Denisse Past{\'{e}}n, Felipe Torres, Benjamín~A. Toledo, Víctor Mu{\~{n}}oz, José Rogan, and Juan~Alejandro Valdivia.
\newblock {Non-universal critical exponents in earthquake complex networks}.
\newblock {\em Physica A: Statistical Mechanics and its Applications}, 491:445--452, 2 2018.
\newblock \href {https://doi.org/10.1016/j.physa.2017.09.064} {\path{doi:10.1016/j.physa.2017.09.064}}.

\bibitem{Pasten2021}
Denisse Past{\'{e}}n.
\newblock {Fractals and complex networks applied to earthquakes}.
\newblock In {\em Basics of Computational Geophysics}, pages 139--151. Elsevier, 2021.
\newblock \href {https://doi.org/10.1016/B978-0-12-820513-6.00021-7} {\path{doi:10.1016/B978-0-12-820513-6.00021-7}}.

\bibitem{Pasten2018b}
Denisse Past{\'{e}}n, Zbigniew Czechowski, and Benjamín Toledo.
\newblock {Time series analysis in earthquake complex networks}.
\newblock {\em Chaos: An Interdisciplinary Journal of Nonlinear Science}, 28(8), 8 2018.
\newblock \href {https://doi.org/10.1063/1.5023923} {\path{doi:10.1063/1.5023923}}.

\bibitem{Pavez2023}
Claudia Pavez-Orrego and Denisse Past{\'{e}}n.
\newblock {Defining the Scale to Build Complex Networks with a 40-Year Norwegian Intraplate Seismicity Dataset}.
\newblock {\em Entropy}, 25(9):1284, 8 2023.
\newblock \href {https://doi.org/10.3390/e25091284} {\path{doi:10.3390/e25091284}}.

\bibitem{Baiesi2005}
M.~Baiesi and M.~Paczuski.
\newblock {Complex networks of earthquakes and aftershocks}.
\newblock {\em Nonlinear Processes in Geophysics}, 12(1):1--11, 1 2005.
\newblock \href {https://doi.org/10.5194/npg-12-1-2005} {\path{doi:10.5194/npg-12-1-2005}}.

\bibitem{Abe2006}
S.~Abe and N.~Suzuki.
\newblock {Complex-network description of seismicity}.
\newblock {\em Nonlinear Processes in Geophysics}, 2006.
\newblock URL: \url{www.nonlin-processes-geophys.net/13/145/2006/}, \href {https://doi.org/10.5194/NPG-13-145-2006} {\path{doi:10.5194/NPG-13-145-2006}}.

\bibitem{Jimenez2013}
Abigail Jim{\'{e}}nez.
\newblock {A complex network model for seismicity based on mutual information}.
\newblock {\em Physica A: Statistical Mechanics and its Applications}, 392(10):2498--2506, 5 2013.
\newblock \href {https://doi.org/10.1016/j.physa.2013.01.062} {\path{doi:10.1016/j.physa.2013.01.062}}.

\bibitem{Nitta2003}
Tohru Nitta.
\newblock {The Computational Power of Complex-Valued Neuron}.
\newblock pages 993--1000. 2003.
\newblock \href {https://doi.org/10.1007/3-540-44989-2{\_}118} {\path{doi:10.1007/3-540-44989-2{\_}118}}.

\bibitem{Pasqualetti2014}
Fabio Pasqualetti, Sandro Zampieri, and Francesco Bullo.
\newblock {Controllability Metrics, Limitations and Algorithms for Complex Networks}.
\newblock {\em IEEE Transactions on Control of Network Systems}, 1(1):40--52, 3 2014.
\newblock \href {https://doi.org/10.1109/TCNS.2014.2310254} {\path{doi:10.1109/TCNS.2014.2310254}}.

\bibitem{Hanrahan2010}
Grady Hanrahan.
\newblock {Computational Neural Networks Driving Complex Analytical Problem Solving}.
\newblock {\em Analytical Chemistry}, 82(11):4307--4313, 6 2010.
\newblock \href {https://doi.org/10.1021/ac902636q} {\path{doi:10.1021/ac902636q}}.

\bibitem{Braha2009}
D.~Braha and Yaneer Bar-Yam.
\newblock {Time-Dependent Complex Networks: Dynamic Centrality, Dynamic Motifs, and Cycles of Social Interactions}.
\newblock pages 39--50. 2009.
\newblock \href {https://doi.org/10.1007/978-3-642-01284-6{\_}3} {\path{doi:10.1007/978-3-642-01284-6{\_}3}}.

\bibitem{Jalili2013}
Mahdi Jalili.
\newblock {Social power and opinion formation in complex networks}.
\newblock {\em Physica A: Statistical Mechanics and its Applications}, 392(4):959--966, 2 2013.
\newblock \href {https://doi.org/10.1016/j.physa.2012.10.013} {\path{doi:10.1016/j.physa.2012.10.013}}.

\bibitem{Klemm2003}
Konstantin Klemm, Víctor~M. Egu{\'{i}}luz, Raúl Toral, and Maxi San~Miguel.
\newblock {Nonequilibrium transitions in complex networks: A model of social interaction}.
\newblock {\em Physical Review E}, 67(2):026120, 2 2003.
\newblock \href {https://doi.org/10.1103/PhysRevE.67.026120} {\path{doi:10.1103/PhysRevE.67.026120}}.

\bibitem{Warrier2021}
Rajeev Warrier.
\newblock {UFC-Fight historical data from 1993 to 2021}, 2021.
\newblock URL: \url{https://www.kaggle.com/datasets/rajeevw/ufcdata}.

\bibitem{Tapology2024}
{Tapology web}.
\newblock {UFC Pay Per View Buys}.
\newblock URL: \url{https://www.tapology.com/search/mma-event-figures/ppv-pay-per-view-buys-buyrate}.

\bibitem{Google2024}
{Google Inc.}
\newblock {Google Trends}, 2024.
\newblock URL: \url{https://trends.google.com/trends/}.

\bibitem{Sizemore2018}
Ann~E. Sizemore and Danielle~S. Bassett.
\newblock {Dynamic graph metrics: Tutorial, toolbox, and tale}.
\newblock {\em NeuroImage}, 180:417--427, 10 2018.
\newblock \href {https://doi.org/10.1016/j.neuroimage.2017.06.081} {\path{doi:10.1016/j.neuroimage.2017.06.081}}.

\bibitem{Rezwan2012}
Rezwan Ahmed and George Karypis.
\newblock {Algorithms for mining the evolution of conserved relational states in dynamic networks}.
\newblock {\em Knowledge and Information Systems}, 33(3):603--630, 12 2012.
\newblock \href {https://doi.org/10.1007/s10115-012-0537-2} {\path{doi:10.1007/s10115-012-0537-2}}.

\bibitem{Zurita2023}
Tomás Zurita-Valencia and Víctor Mu{\~{n}}oz.
\newblock {Characterizing the Solar Activity Using the Visibility Graph Method}.
\newblock {\em Entropy}, 25(2):342, 2 2023.
\newblock \href {https://doi.org/10.3390/e25020342} {\path{doi:10.3390/e25020342}}.

\bibitem{Abe2004}
S~Abe and N~Suzuki.
\newblock {Scale-free network of earthquakes}.
\newblock {\em Europhysics Letters (EPL)}, 65(4):581--586, 2 2004.
\newblock \href {https://doi.org/10.1209/epl/i2003-10108-1} {\path{doi:10.1209/epl/i2003-10108-1}}.

\bibitem{Abe2011}
Sumiyoshi Abe, Denisse Past{\'{e}}n, Víctor Mu{\~{n}}oz, and Norikazu Suzuki.
\newblock {Universalities of earthquake-network characteristics}.
\newblock {\em Chinese Science Bulletin}, 2011.
\newblock URL: \url{https://link.springer.com/article/10.1007/s11434-011-4767-6}, \href {https://doi.org/10.1007/S11434-011-4767-6} {\path{doi:10.1007/S11434-011-4767-6}}.

\bibitem{Zou2014}
Yong Zou, Michael Small, Zonghua Liu, and Jürgen Kurths.
\newblock {Complex network approach to characterize the statistical features of the sunspot series}.
\newblock {\em New Journal of Physics}, 2014.
\newblock URL: \url{https://iopscience.iop.org/article/10.1088/1367-2630/16/1/013051 https://iopscience.iop.org/article/10.1088/1367-2630/16/1/013051/meta}, \href {https://doi.org/10.1088/1367-2630/16/1/013051} {\path{doi:10.1088/1367-2630/16/1/013051}}.

\bibitem{Pasten2017}
Denisse Past{\'{e}}n, Felipe Torres, Benjamín Toledo, Víctor Mu{\~{n}}oz, José Rogan, and Juan~Alejandro Valdivia.
\newblock {Time-Based Network Analysis Before and After the 8.3 Illapel Earthquake 2015 Chile}.
\newblock {\em The Chile-2015 (Illapel) Earthquake and Tsunami}, 2017.
\newblock URL: \url{https://link.springer.com/chapter/10.1007/978-3-319-57822-4_10}, \href {https://doi.org/10.1007/978-3-319-57822-4{\_}10} {\path{doi:10.1007/978-3-319-57822-4{\_}10}}.

\bibitem{Hu2008}
Hai-Bo Hu and Xiao-Fan Wang.
\newblock {Unified index to quantifying heterogeneity of complex networks}.
\newblock {\em Physica A: Statistical Mechanics and its Applications}, 387(14):3769--3780, 6 2008.
\newblock \href {https://doi.org/10.1016/j.physa.2008.01.113} {\path{doi:10.1016/j.physa.2008.01.113}}.

\bibitem{Newman2001}
M.~E.~J. Newman.
\newblock {The structure of scientific collaboration networks}.
\newblock {\em Proceedings of the National Academy of Sciences}, 98(2):404--409, 1 2001.
\newblock \href {https://doi.org/10.1073/pnas.98.2.404} {\path{doi:10.1073/pnas.98.2.404}}.

\bibitem{Barabasi1999}
Albert~László Barab{\'{a}}si and Réka Albert.
\newblock {Emergence of scaling in random networks}.
\newblock {\em Science}, 1999.
\newblock URL: \url{https://www.science.org/doi/10.1126/science.286.5439.509}, \href {https://doi.org/10.1126/science.286.5439.509} {\path{doi:10.1126/science.286.5439.509}}.

\bibitem{Freedman2005}
David Freedman.
\newblock {\em {Statistical Models: Theory and Practice}}.
\newblock Cambridge University Press, Cambridge, 2005.
\newblock \href {https://doi.org/10.1017/CBO9781139165495} {\path{doi:10.1017/CBO9781139165495}}.

\bibitem{Schober2018}
Patrick Schober, Christa Boer, and Lothar~A. Schwarte.
\newblock {Correlation Coefficients: Appropriate Use and Interpretation}.
\newblock {\em Anesthesia {\&} Analgesia}, 126(5):1763--1768, 5 2018.
\newblock \href {https://doi.org/10.1213/ANE.0000000000002864} {\path{doi:10.1213/ANE.0000000000002864}}.

\bibitem{Duch2010}
Jordi Duch, Joshua~S. Waitzman, and Luís A.~Nunes Amaral.
\newblock {Quantifying the Performance of Individual Players in a Team Activity}.
\newblock {\em PLoS ONE}, 5(6):e10937, 6 2010.
\newblock \href {https://doi.org/10.1371/journal.pone.0010937} {\path{doi:10.1371/journal.pone.0010937}}.

\bibitem{Fletcher2018}
Sam Fletcher and Md~Zahidul Islam.
\newblock {Comparing sets of patterns with the Jaccard index}.
\newblock {\em Australasian Journal of Information Systems}, 22, 3 2018.
\newblock \href {https://doi.org/10.3127/ajis.v22i0.1538} {\path{doi:10.3127/ajis.v22i0.1538}}.

\end{thebibliography}

\end{document}